\documentclass[12pt, preprint,numberedappendix,twocolappendix]{emulateapj}

\newcommand\bibinc{n}		

\usepackage{ulem}
\usepackage{subeqnarray}
\usepackage{amsmath}
\usepackage{hyperref}
\hypersetup{colorlinks=true,citecolor=blue}

\DeclareMathSymbol{\varOmega}{\mathord}{letters}{"0A}
\DeclareMathSymbol{\varSigma}{\mathord}{letters}{"06}
\DeclareMathSymbol{\varPsi}{\mathord}{letters}{"09}

\newcommand{\Eq}[1]{Equation\,(\ref{#1})}

\newcommand{\Sec}[1]{Section~\ref{#1}}

\newcommand{\Fig}[1]{Figure~\ref{#1}}

\newcommand {\mesa} {{\tt MESA}}

\begin{document}

\slugcomment{Accepted at ApJ}

\shorttitle{Internally Heated Hot Jupiters}
\shortauthors{T.D. Komacek \& A.N. Youdin}

\title{Structure and Evolution of Internally Heated Hot Jupiters }
\author{Thaddeus D. Komacek$^1$ and Andrew N. Youdin$^2$} \affil{$^1$Department of Planetary Sciences and Lunar and Planetary Laboratory
 $^2$Department of Astronomy and Steward Observatory \\
 University of Arizona, Tucson, AZ, 85721 \\
\url{tkomacek@lpl.arizona.edu}} 
\begin{abstract}
Hot Jupiters receive strong stellar irradiation, producing equilibrium temperatures of $1000 - 2500 \ \mathrm{Kelvin}$.    Incoming irradiation directly heats just their thin outer layer, down to pressures of $\sim 0.1$ bars. In standard irradiated evolution models of hot Jupiters, predicted transit radii are too small.   
Previous studies have shown that deeper heating -- at a small fraction of the heating rate from irradiation --  can explain observed radii.  Here we present a suite of evolution models for HD 209458b where we systematically vary both the depth and intensity of internal heating, without specifying the uncertain heating mechanism(s).  Our models start with a hot, high entropy planet whose radius decreases as the convective interior cools. The applied heating suppresses this cooling.
We find that very shallow heating -- at pressures of $1 - 10 \ \mathrm{bars}$ -- does not significantly suppress cooling, unless the total heating rate is $\gtrsim 10\%$ of the incident stellar power.  Deeper heating, at $100$ bars, requires heating at only $1\%$ of the stellar irradiation to explain the observed transit radius of $1.4 R_{\rm Jup}$ after 5 Gyr of cooling.  
In general, more intense and deeper heating results in larger hot Jupiter radii.  
Surprisingly, we find that heat deposited at $10^4 \ \mathrm{bars}$ -- which is exterior to $\approx 99\%$ of the planet's mass -- suppresses planetary cooling as effectively as heating at the center.  In summary, we find that relatively shallow heating is required to explain the radii of most hot Jupiters, provided that this heat is applied early and persists throughout their evolution.
\end{abstract}
\section{Introduction}
\begin{figure}
	\centering
	\includegraphics[width=0.5\textwidth]{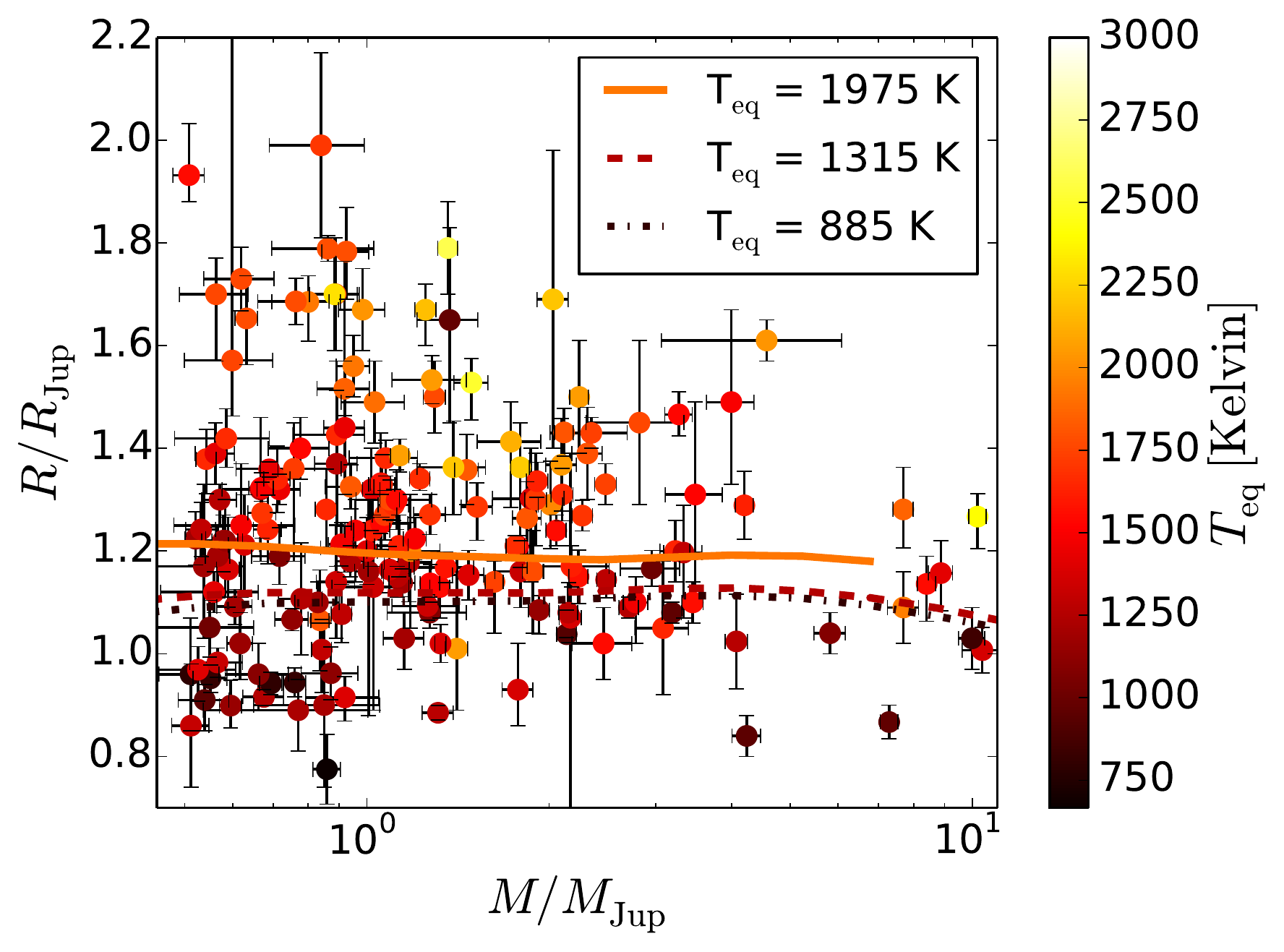}
	\caption{Masses and radii of known hot Jupiters, relative to Jupiter and colored by equilibrium temperature, $T_{\mathrm{eq}} = [F_{\star}/(4\sigma)]^{1/4}$, where $F_{\star}$ is the incident stellar flux and $\sigma$ the Stefan-Boltzmann constant. The over-plotted lines show \cite{Fortney:2007ta} model radii after 3.2 Gyr of cooling with irradiation levels set by the $T_{\rm eq}$ values as labelled and with no solid core. Simple irradiated cooling models cannot produce the observed range of radii. Data is taken from the \url{http://exoplanets.org/} database \citep{Han:2014}. } 
	\label{fig:MRhotjups}
\end{figure}

Since the first transit detections of an extrasolar planet \citep{Charbonneau_2000,Henry:2000}, the close-in extrasolar giant planet, or ``hot Jupiter,'' population has proved to be enigmatic. It was recognized early that many of these hot Jupiters have radii larger than expected from standard models considering only cooling from a high-entropy initial state (for reviews see \citealp{fortney_2009,Baraffe:2010xe,Baraffe:2014,Laughlin:2015}). A mass-radius diagram of the hot Jupiter sample is shown in \Fig{fig:MRhotjups}, with cooling models from \cite{Fortney:2007ta} assuming varying amounts of incident stellar flux over-plotted. Approximately one-half of the observed hot Jupiters have radii above expectations from evolutionary models. Though including irradiation in these models causes a deep radiative zone which pushes the radiative-convective boundary (RCB) to higher pressures and thereby reduces the planetary cooling rate, this only affects radii by $\lesssim 20\%$ \citep{Guillot:1996,Arras:2006kl,Fortney:2007ta}. \\
\indent There is also a trend of increasing planet radius with increasing equilibrium temperature, found by \cite{Laughlin_2011} and visible by eye in \Fig{fig:MRhotjups}. Further analysis shows that planets with equilibrium temperatures $T_{\mathrm{eq}} \lesssim 1000 \ \mathrm{Kelvin}$ have radii which match the expectations of purely cooling models \citep{Demory:2011,Miller:2011}. Re-inflated hot Jupiters have been proposed as potential additional evidence that stellar irradiation is what drives radius inflation \citep{Lopez:2015}, and there is a growing observed population of re-inflated hot Jupiters around post-main sequence stars \citep{Grunblatt2016,Hartman2016}. Hence, the mechanisms responsible for enlarged hot Jupiters are somehow correlated with incident stellar flux. \\
\indent There are three classes of explanations for the radius anomaly of hot Jupiters: tidal mechanisms, modifications to our understanding of the microphysics of hot Jupiters,
and incident stellar-flux driven mechanisms \citep{Weiss:2013,Baraffe:2014}. Tidal dissipation was the first proposed mechanism for explaining the bloated radii of many hot Jupiters \citep{Bodenheimer:2001}, and as such has been followed up by a variety of studies \citep{Jackson:681,Ibgui:2009,Miller:2009,Ibgui:2010,Leconte:2010a}. However, tidal dissipation nominally requires the eccentricity of the planet to be pumped up by an external companion, as tidal dissipation damps eccentricity. \cite{Arras:2010} proposed that thermal tides in the atmosphere of the planet itself can torque it away from synchronous rotation, enabling the interior of the planet to couple to the stellar gravitational tidal force and cause dissipation, increasing the viability of this mechanism. The second class of mechanisms, those which do not require internal heating, includes enhanced opacities \citep{Burrows:2007bs} and lowered internal heat transport due to double-diffusive convection \citep{Chabrier:2007,Leconte:2012}. Though certainly important for understanding the internal structure of gas giants, it is unclear if these microphysical mechanisms to increase the radius of hot Jupiters scale with the incoming stellar flux. \\
\indent The third class of mechanism involves transport of a fraction of the heat in the high-entropy irradiated region of the planet downward to the interior, where it then dissipates and modifies the entropy of the internal adiabat. These mechanisms are all necessarily linked to the vigorous atmospheric circulation in hot Jupiter atmospheres, where $\sim\mathrm{km/s}$ east-west winds are driven by the large day-to-night temperature contrasts in these atmospheres \citep{showman_2002,Cooper:2005,Menou:2009,Showmanetal_2009,Showman_2009,Rauscher:2010,Showman_Polvani_2011,perna_2012,Mayne:2014,Kataria2016,Komacek:2015,Komacek:2017}. This class of mechanisms can additionally be split into two classes: the hydrodynamic and magnetohydrodynamic mechanisms. The latter, known as ``Ohmic dissipation,'' utilizes currents driven in the partially ionized atmosphere (threaded by a dipolar planetary magnetic field) which then resistively dissipate in the interior of the planet. This mechanism was proposed by \cite{Batygin_2010}, with a variety of follow-up studies \citep{Perna_2010_2,Batygin_2011,Heng:2012,Huang_2012,Menou:2012fu,Menou_2012,Rauscher_2013,Wu:2013,Rogers:2014,Rogers:2020,Ginzburg:2015a} performed either supporting or refuting to varying degrees the Ohmic dissipation hypothesis. Models of purely kinematic magnetohydrodynamics normally attain dissipation rates large enough to explain radius anomalies, but self-consistent magnetohydrodynamic simulations do not. \\
\indent Even for nominally simpler hydrodynamic dissipation mechanisms, several details remain poorly understood. The first of this class of mechanism proposed was downward kinetic energy transport \citep{Guillot_2002,showman_2002}. In this picture, the $\sim 10-100 \ \mathrm{m/s}$ vertical winds in hot Jupiter atmospheres transport energy downward. This energy is then dissipated in shear layers near the RCB through, for example, Kelvin-Helmholtz instabilities. Another possible mechanism, the ``Mechanical Greenhouse'' \citep{Youdin_2010}, involves downward transport of heat by forced turbulent mixing in the outer radiative zone of the planet. Building upon this, \cite{Tremblin:2017} showed using a two-dimensional steady-state dynamical model that the large-scale circulation itself can produce enough downward entropy transport to explain the radius of HD 209458b. These are viable mechanisms to explain the hot Jupiter radius anomaly. However, no deep time-dependent simulations of hot Jupiter atmospheres have yet been performed to quantify the three-dimensional atmospheric circulation at levels where dissipation would strongly affect the interior entropy. \\
Though there have been a number of mechanisms proposed to explain the hot Jupiter radius anomaly, none of those discussed above have been shown to explain the entire sample of hot Jupiters. As a result, we focus in this work not on specific mechanisms but on how internal heating of any strength at any location affects the evolution and structure of hot Jupiters. \cite{Spiegel:2013} showed that adding heat in the atmosphere of a hot Jupiter was more efficient at increasing pressure levels (see their Figure 4), with a markedly nonlinear relationship between transit radius and pressure of dissipation. They also showed that heating in the center of the planet enables a long-timescale radius equilibrium to be reached, which is not possible when heating is applied in the atmosphere alone. However, they did not examine the impact of heating at a given pressure level on the structural evolution of the hot Jupiter, as their numerical model decoupled the evolution of the radiative atmosphere and convective interior of the planet. As a result, the relationship between radius and deposited power from \cite{Spiegel:2013} is determined only for jointly specified surface gravity and effective temperature, even though both depend on the deposited power itself. \\
\indent It is possible, as has been shown by the self-similar analytic model of \cite{Ginzburg:2015}, that outer convection zones can be forced due to strong heating. Their analytic model shows that the presence or absence of an outer convective zone is determined by a comparison between the combination of the strength and depth of deposited heating and the incoming stellar flux. The analytic model of \cite{Ginzburg:2015} also predicts the transition, in terms of heating strength and depth, where heating in the outer radiative zone begins to have a substantial effect on the resulting transit radius. In this work we test these predictions using a detailed numerical evolutionary model, which also enables us to consider the regime where heating occurs deeper than the inner radiative-convective boundary. This regime is not directly considered in \cite{Ginzburg:2015}. Another possibility which we take into account is that dissipation occurs at a given structural location within the planet, for example the radiative-convective boundary, which is not fixed in pressure with time.  \\ 
\indent Here we use the stellar \& planetary structure code \mesa \ \citep{Paxton:2011,Paxton:2013,Paxton:2015} to compute the evolution of a hot Jupiter with different amounts of heating deposited at different depths or structural locations. These models allow a detailed understanding of the structural and evolutionary pathways of planets subject to different types of heating. This paper is organized as follows. \Sec{sec:Methods} describes the numerical setup used, including choices for heating profiles. \Sec{sec:cooling} describes the general theoretical framework that we use to  interpret our results. \Sec{sec:results} displays our results for how heating of varying strength and deposition depth affects the structure and evolution of hot Jupiters. We compare these results to the analytic theory of \cite{Ginzburg:2015} in \Sec{sec:discussion}, and delineate conclusions in \Sec{sec:conclusions}.
\section{Methods}
\label{sec:Methods}
\subsection{Planetary Structure Equations}
\label{sec:structure}
We use \mesa \ to solve the following equations of ``stellar'' structure \citep{Chandrasekhar:1939,Kippenhahn:2012}, applied to giant planets:
\begin{equation}
\label{eq:mass}
\frac{dm}{dr} = 4\pi r^2 \rho\mathrm{,}
\end{equation}
\begin{equation}
\label{eq:hydro}
\frac{dP}{dm} = -\frac{G m}{4 \pi r^4}\mathrm{,}
\end{equation}
\begin{equation}
\label{eq:heat}
\frac{dL}{dm} = \frac{d\left(L_{\mathrm{rad}} + L_{\mathrm{conv}}\right)}{dm} = \epsilon_{\mathrm{grav}} + \epsilon_{\mathrm{extra}}	\mathrm{,}
\end{equation}
\begin{equation}
\label{eq:transport}
\frac{dT}{dm} = -\frac{GmT}{4\pi r^4 P}\nabla\mathrm{.}
\end{equation}
\Eq{eq:mass} is the mass conservation equation giving the enclosed mass, $m$, at radius, $r$, with mass density $\rho$.
\Eq{eq:hydro} expresses hydrostatic equilibrium of the pressure, $P$, with the gravitational constant, $G$. \Eq{eq:heat} states energy conservation, where the outgoing luminosity, $L$, computed by \mesa \ includes both radiative ($L_{\mathrm{rad}}$) and (where convectively unstable) convective  ($L_{\mathrm{conv}}$) components.  The relevant energy sources include the cooling term, $\epsilon_{\mathrm{grav}} = -T dS/dt$, i.e.\ the loss of entropy, $S$, that drives gravitational Kelvin-Helmholz contraction, and $\epsilon_{\mathrm{extra}}$, the ``extra'' energy deposition described in detail below. \Eq{eq:transport} is the energy transport equation, in terms of the logarithmic gradient, $\nabla \equiv d\mathrm{ln}T/d\mathrm{ln}P$, where $T$ is temperature.  \mesa\ uses mixing length theory to calculate $\nabla = \nabla_{\mathrm{rad}}$, the radiative gradient, (or $\nabla \simeq \nabla_{\mathrm{ad}}$, the convective gradient) in radiative (or convective) regions (respectively). Specifically, the Schwarzchild criterion sets the temperature gradient $\nabla = d\mathrm{ln}T/d\mathrm{ln}P$ to the smaller of the adiabatic gradient $\nabla_{\mathrm{ad}}$ or the radiative gradient
\begin{equation}
\label{eq:radgrad}
\nabla_{\mathrm{rad}} = \frac{3}{64\pi \sigma G}\frac{\kappa LP}{mT^4} \mathrm{.}
\vspace{0.01cm}
\end{equation}
In \Eq{eq:radgrad}, $\sigma$ is the Stefan-Boltzmann constant and $\kappa$ is the local opacity. The set of Equations (\ref{eq:mass}) - (\ref{eq:transport}) is closed with a thermodynamic equation of state \citep{Saumon:1995,Paxton:2013}, and with tabulated opacities, needed for $\nabla_{\rm rad}$, that assume a dust-free Solar composition  \citep{Freedman:2008,Paxton:2011,Paxton:2013}. 
\subsection{Numerical Model}
\subsubsection{Setup}
\indent The initial models for our \mesa \ evolutionary calculations are computed as described in \cite{Paxton:2013}. We choose a mass relevant for HD 209458b and assume an initial radius of $2.3 \ R_{\mathrm{Jup}}$ to compute an adiabatic starting planet model. The long-term evolution, at times $\gtrsim 10 \ \mathrm{Myr}$, is independent of the initial radius and hence is also independent of the initial entropy \citep{Arras:2006kl}. However, note that the Kelvin-Helmholtz contraction timescale of a planet is inversely related to both its radius and luminosity. This means that if we instead chose a larger starting radius (and hence entropy) the planet would have a more rapid early evolution but after this initial cooling phase would end up on the same evolutionary track as a planet with a smaller initial radius.
For comparison, the present-day transit radius of HD 209458b is $\approx 1.35 \ R_\mathrm{Jup}$. We evolve the initial model with external irradiation and internal heating, as described below. \\
\indent Equations (\ref{eq:mass})-(\ref{eq:transport}) are solved using the Henyey method \citep{Henyey:1959,Bodenheimer:2007,Kippenhahn:2012}, including automatic mesh refinement \citep{Paxton:2011}.   The outer boundary pressure and temperature are fixed at the location where the optical depth $\tau$ to outgoing radiation is $2/3$. We ensure that the Henyey residuals are small enough to not affect structure for all of our evolution calculations. We describe the specified external irradiation in Section \ref{sec:mesacap}, our choices for $\epsilon_{\mathrm{extra}}$ in Section \ref{sec:heatsetup}, and our parameter choices in Section \ref{sec:params}.
\subsubsection{Irradiation}
\label{sec:mesacap}
\indent The irradiation of the exoplanet is included as an energy generation rate $\epsilon_{\mathrm{extra}} = F_{\star}/(4\Sigma_p)$, which is applied where the outer mass column $\Sigma$  is less than the chosen $\Sigma_p$ (see \Sec{sec:params} for specific parameter values).
With $F_{\star}$ as the incoming stellar flux, this heating drives a heat flux of $F_{\star}/4$, equivalent to the average of the incoming radiation over the planet's spherical surface.
This is the $F_{\star}-\Sigma_p$ irradiation routine \citep{Paxton:2013}, which has also been used by \cite{Owen:2015,Valsecchi:2015}. This irradiation method agrees with the detailed ``grey irradiated" solutions of \citet{Guillot:2010}, as shown  
in \cite{Paxton:2013} (see their Figure 3, which shows agreement within $\sim 1-2\%$ after 100 Myr of evolution). 
\subsubsection{Internal Heating}
\label{sec:heatsetup}
In this work, we consider the impacts of heating at different depths and structural locations within a given hot Jupiter on the resulting evolution of the planet. Though this heating can be interpreted as due to dynamical processes which deposit heat from near-photospheric levels to greater depths, we do not consider the impacts of specific heating mechanisms. Instead, we add an extra dissipation $\epsilon_{\mathrm{extra}}$ (computed each time step) that is taken to be a Gaussian with standard deviation $0.5 H$, where $H = p/(\rho g)$ is a pressure scale height. 
This heating profile is similar to that applied by \cite{Spiegel:2013}, but as we are using a global planetary structure code (instead of a detailed atmosphere model matched onto an adiabat) we can calculate the impact of heating near or below the RCB. \\ 
\indent We consider heating deposited at pressures of $P_{\mathrm{dep}} = 1 - 10^4 \ \mathrm{bars}$ (distributed as described above), at the center of the planet, and at the
inner RCB. The integrated heating rates,
\begin{equation}
\Gamma = \int_0^M \epsilon_{\mathrm{extra}} dm \mathrm{,}
\end{equation}
are set to different fractions of the irradiation as 
\begin{equation}
\gamma \equiv \Gamma/L_{\mathrm{irr}} = 10^{-5} - 0.1
\end{equation}
for this grid of simulations. Heating rates are written as a fraction of the incident stellar power $L_{\mathrm{irr}} = 2.4 \times 10^{29} \ \mathrm{erg} \ \mathrm{sec}^{-1}$ relevant for HD 209458b. \\
\indent We choose an upper limit of $\gamma = 0.1$ because a $100\%$ conversion of starlight to kinetic energy, which is transported to depth via either hydrodynamic or magnetohydrodynamic mechanisms, is unrealistic. Thus we choose $10\%$ conversion as an upper limit, consistent with previous studies. In the case of hydrodynamic dissipation, the upper end of these values for the normalized heating rate $\gamma$ can be motivated from the fraction of atmospheric kinetic energy dissipated to heat, which is unknown for hot Jupiters but is expected to be $\sim 1- 10\%$ based on Earth \citep{Peixoto:1992,Guillot_2002,Schubert:2013}. It is expected that for Ohmic dissipation $\gamma \lesssim 1-5 \%$ \citep{Perna_2010_2,Batygin_2011,Rauscher_2013,Ginzburg:2015a} and may be up to two orders of magnitude smaller \citep{Rogers:2020,Rogers:2014}, and for tidal dissipation $\gamma \lesssim 10\%$ \citep{Arras:2010} and may be as small as $\sim 0.1\%$ \citep{Bodenheimer:2001,Jackson:681}. When we apply heating at the RCB, $P_{\mathrm{dep}} = P_{\mathrm{RCB}}$ follows the location of the boundary. Heating at this boundary can be motivated by either downward mixing of heat by large-scale eddies or communication between atmospheric motions and the deep interior resulting in shear instabilities at this interface \citep{Guillot_2002,showman_2002,Youdin_2010}. 
\subsubsection{Parameter Choices}
\label{sec:params}
\indent We keep all external planetary parameters constant, running a suite of simulations varying only the integrated heating rates $\Gamma$ and deposition pressure $P_{\mathrm{dep}}$. We use a mass ($0.69 \ M_{\mathrm{Jup}}$), composition ($Y = 0.24, Z = 0.02$), and irradiation flux relevant for HD 209458b, taken from \cite{Guillot_2002}. As in \cite{Guillot_2002}, our model does not include a heavy element core. As a result, our resulting radii are upper limits for the assumed heavy element composition. \\
\indent 
For our irradiation routine, we choose $\Sigma_p = 250 \ \mathrm{g} \ \mathrm{cm}^{-2}$, corresponding to an opacity $\kappa_{\mathrm{vis}} = 4 \times 10^{-3} \ \mathrm{cm}^2 \ \mathrm{g}^{-1}$  to incoming radiation, as in
 \cite{Fortney:2008,Guillot:2010,Owen:2015}. 
We use $F_{\star} = 1.0012 \times 10^9 \ \mathrm{erg} \ \mathrm{cm}^{-2} \ \mathrm{s}^{-1} = 1200 F_{\oplus}$ (where $F_{\oplus}$ is Earth's incident flux), which is equivalent to an equilibrium temperature $T_{\mathrm{eq}} = 1450 \ \mathrm{Kelvin}$ assuming full longitudinal redistribution of heat. We evolve our modeled planets to 5 Gyr, a typical age for main sequence systems and a nominal stopping point to compare with observed transit radii (e.g. \citealp{Huang_2012,Wu:2013}).
\section{Cooling Regimes of Hot Jupiters}
\label{sec:cooling}
\begin{figure*}
	\centering
	\includegraphics[width=1\textwidth]{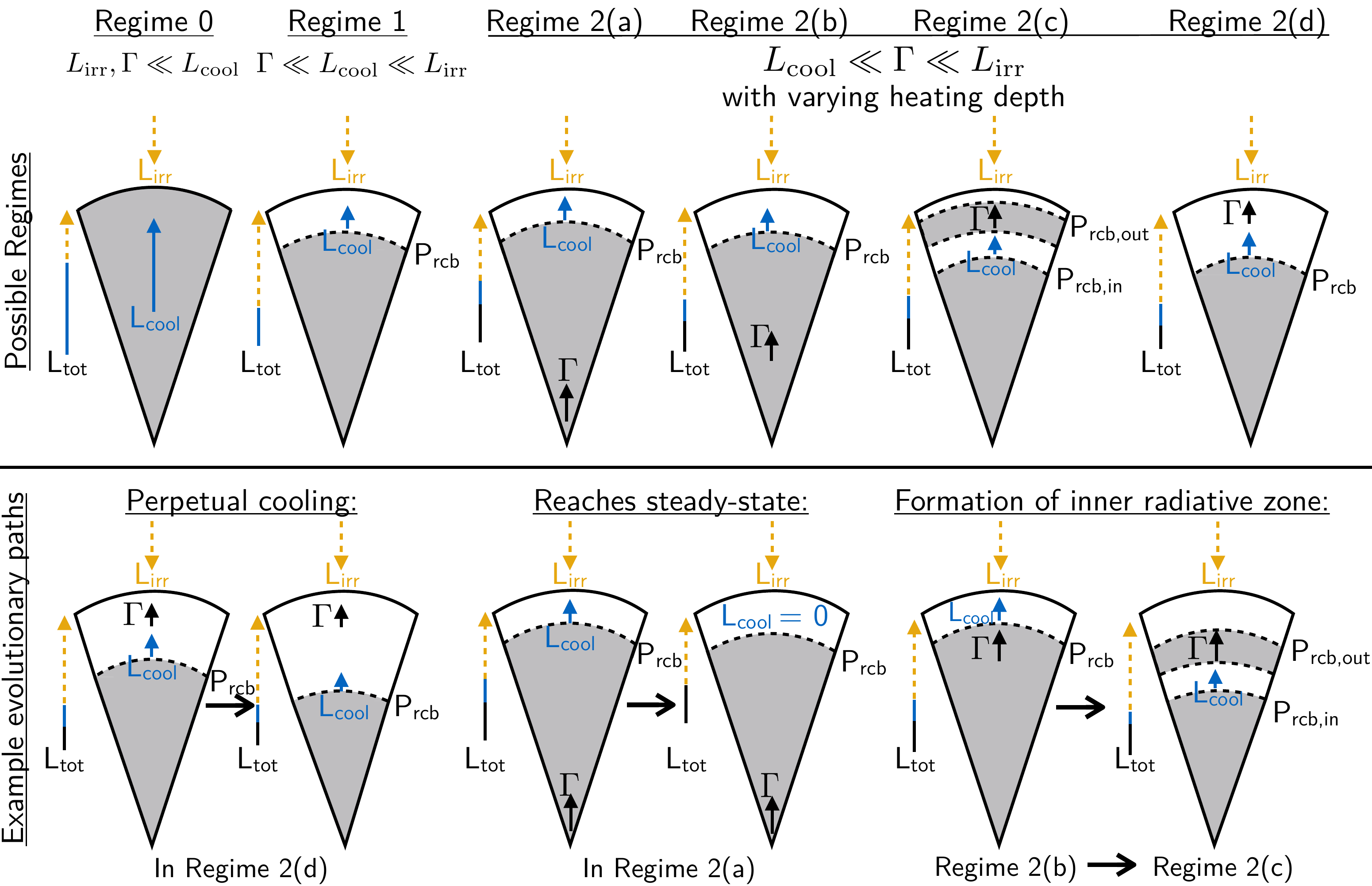}
	\caption{Schematic pie slices of the internal structure of a hot Jupiter at different evolutionary stages and subject to different strengths and depths of heating. The top panels identify the key regimes: photospheric cooling (regime 0), irradiated cooling (regime 1), and irradiated cooling with internal heating (regime 2). The bottom panels illustrate different evolutionary paths within regime 2. Shaded regions are convective, and white regions are radiative. The  strengths of irradiation, the cooling luminosity, and heat deposition are depicted as arrows, which are not drawn to scale. The arrow on the left-hand-side of each schematic shows the contribution (not to scale) of each source of these to the total outgoing luminosity. See \Sec{sec:cooling} for discussion.}
	\label{fig:cartoon}
\end{figure*}
\subsection{Definition of Regimes}
We here describe the main cooling regimes that occur in the evolution of giant gaseous planets, subject to different levels of irradiation and different amounts and depths of internal heating. Understanding these regimes and the evolution between them facilitates interpretation of our results in \Sec{sec:results}. \Fig{fig:cartoon} shows a schematic (which will be referred to throughout this section) of the various possible structural regimes a gas giant may lie in (top panels), along with example evolutionary paths between these regimes (bottom panels). \\ 
\indent First, consider regime 0, which applies for planets which have a cooling luminosity that exceeds any external irradiation (i.e. the cooling luminosity $L_{\mathrm{cool}} \gg$ the irradiation power $L_{\mathrm{irr}}$).  Such a planet has a fully convective interior and radiates from a RCB at the planet's photosphere.  Regime 0 applies to weakly irradiated planets or to irradiated planets with high initial entropies.  Our strongly irradiated models are never in regime 0, and if we were to start with higher entropies this phase would be very brief.

As an irradiated regime 0 planet cools and $L_{\mathrm{cool}}$ declines, it eventually enters regime 1, where external irradiation exceeds the planet's cooling luminosity (i.e. $L_{\mathrm{irr}} \gg L_{\mathrm{cool}}$).  For regime 1, we also require that the cooling exceeds any extra heat that is input from e.g. turbulent, Ohmic or tidal dissipation.  Thus in regime 1, this extra heat does not yet significantly affect evolution.  As the planet cools in regime 1, the outer radiative zone recedes below the photosphere, i.e. the RCB increases in pressure with time as the entropy of the convective interior decreases, because the majority of planetary cooling is from the convective interior \citep{Arras:2006kl}.  Planets without any extra heat will remain in regime 1.  Between the convective interior and radiative exterior, intermediate convective zones and radiative windows may develop due to abrupt changes in the EOS and/or opacity. We ignore this detail for now, by noting that the key issue is the cooling rate from the innermost RCB, which connects to the central entropy. We will shortly address intermediate convective zones triggered by extra heating, which is not relevant in regime 1.  


Regime 2, where the planet's extra heating exceeds the cooling rate (i.e. the deposited power $\Gamma \gg L_{\mathrm{cool}}$), is the most relevant for our study.  The transition from regime 1 to 2 occurs naturally due to the gradual decline in $L_{\rm cool}$ with time as the internal entropy decreases.  During regime 2, the extra heating reduces the cooling luminosity as either luminosity ``replacement" or luminosity ``suppression.'' To try to be exhaustive, we describe four sub-regimes, which we label:
\begin{itemize}
\item 2(a): The limiting case of heating at the very center of the planet, or at the boundary between a solid core and the gaseous envelope.
\item 2(b): Heating in the convective interior at an intermediate radius.
\item 2(c): Heating outside the convective interior which triggers an outer convective zone above -- and a corresponding radiative window below -- the heating level.
\item 2(d): Heating outside the convective interior in a fully radiative zone.
\end{itemize}

Regime 2(a) represents the simplest case of heating at the deepest possible level of the interior.  Here the added heat simply supplies, or ``replaces,'' some of the planet's luminosity as $L_{\rm cool} = L_{\rm RCB} - \Gamma$.  Only hot Jupiters in regime 2(a) can reach an exact steady state with $L_{\rm cool} = 0$ within stellar main sequence lifetimes.  However, planets in other regimes can (and do) have a cooling time that is longer than the age of the system, with negligible entropy loss.

Regime 2(b) is similar to 2(a) except heating is no longer exactly at the center. The heating is still within the convective interior and can still be thought of as replacing some of the RCB luminosity to decrease cooling. In this regime, the convective region below the heating layer emits cooling luminosity at a finite (but perhaps negligible) rate. A lower bound on the initial cooling luminosity is set by $L_\mathrm{rad,ad}(P_\mathrm{dep})$ which means the radiative luminosity along the adiabat at the location of applied heating. \footnote{This estimate assumes a single depth of heat deposition, and could be refined for more broadly distributed heating. However, this refinement is not needed for a basic understanding.} This minimum initial amount of cooling can be quite low for large $P_\mathrm{dep}$.


In regimes 2(c) and 2(d), the heating is outside the convective interior.  In both regimes, heating pushes the RCB of the convective interior deeper in the planet, which suppresses cooling.  The difference between the regimes is that in 2(c) the heating is deep and intense enough to trigger convection above -- and a radiative window below -- where heat is deposited.  In regime 2(c), the outer boundary of the convective interior (below the radiative window) is pushed to greater depths, which significantly reduces cooling compared to regime 2(d). \cite{Ginzburg:2015} analytically derived the conditions for triggering an outer convective zone, explaining the resulting strong reduction in planetary cooling.
\subsubsection{Comparison with the Evolutionary Stages of Ginzburg \& Sari (2016)}
Before discussing the evolution between regimes, it is useful to compare our framework to the analysis of \cite{Ginzburg:2015a}, who considered the stages that a planet with applied internal heating evolves through (see their Appendix A and Figure 7). Note that \cite{Ginzburg:2015a} consider a different heating profile: a power-law in optical depth with a cutoff above a specified optical depth value. However, a comparison is still possible as the steepness of their power-law nicely corresponds to different cases of deep vs. shallow heating that we study here. \\
\indent Stage 1 of \cite{Ginzburg:2015a} corresponds to our regime 0 of a fully convective planet that cools independent of heating or irradiation. For hot Jupiters, we again emphasize that this phase is either non-existent or very brief, only occurring in the initial evolution.  Stage 2 of \cite{Ginzburg:2015a} is the same as our regime 1.  We define this regime in terms of heating and cooling rates, and \cite{Ginzburg:2015a} express the limits equivalently in terms of optical depths in their self-similar model.  Stage 3 of \cite{Ginzburg:2015a} corresponds to our regimes 2(c) and 2(d), which in turn corresponds to the case I and II (respectively) that \cite{Ginzburg:2015a} describe for the depth of heating.   Our regime 2(a) corresponds to the special case of central heating (in case I of \citealp{Ginzburg:2015a}) in either stage 3 (while cooling is proceeding) or stage 4 (once steady state is reached) of \cite{Ginzburg:2015a}. \\
\indent Along with all these similarities, there are two subtle but notable differences.  We do not describe the general case of their stage 4, a steady state in which the deep interior is fully radiative and isothermal, all the way to the center.  The overlap with the limiting case of stage 4 mentioned above is for heating that occurs precisely at the center.  We do not describe the general case of stage 4 in \cite{Ginzburg:2015a} simply because main sequence stellar lifetimes are not long enough for hot Jupiters to reach this stage.  Specifically we find that even with deep heating, a radiative window may never open (see e.g. our $\gamma = 10^{-2}$ heating efficiency at $P_{\rm dep} = 10^4$ bars case in \Sec{sec:results}) or never extends close to the center (e.g. our $\gamma= 10^{-3}$ at $P_{\rm dep} = 10^4$ bars case). This is because by the time the radiative window opens (if it does), the deep heating has reduced the cooling luminosity to very low values. As a result, the Kelvin-Helmholtz timescale is much longer than main sequence lifetimes, and this timescale steadily increases as the radiative window deepens. \\
\indent The second difference is that \cite{Ginzburg:2015a} do not explicitly describe our regime 2(b), during which the internal heating is both significant and within the convective interior.  Our regime 2(b) is distinct from the closest related stages of \cite{Ginzburg:2015a}; internal heating is not relevant in their stage 2 and heating is disconnected from the convective interior in their stage 3.  We find that regime 2(b) is significant for understanding hot Jupiter evolution.  For one thing, some planets with deep heating will spend their entire late stage cooling (i.e. after regime 1, which is independent of internal heating) in regime 2(b), never opening a radiative window.  Second, for planets that do evolve from regime 2(b) to 2(c) by opening a radiative window, the suppression of cooling is much more complete during regime 2(b).  In these cases, the final radius correlates positively with the (logarithmic) fraction of time spent in regime 2(b).  Our numerical results will more clearly show the importance of regime 2(b) in interesting regions of parameter space.  To our knowledge, a simple analytic theory that includes the equivalent of our regime 2(b) has not yet been published.


\subsection{Evolution}
To better understand these regimes, it helps to consider how a planet enters regime 2, which is when the heating first becomes significant. A crucial issue is the depth that characterizes heat deposition, $P_\mathrm{dep}$, relative to the RCB depth, $P_\mathrm{RCB}$, at this moment. With ``deep heating'' $P_\mathrm{dep} > P_\mathrm{RCB}$ and regime 2 begins as 2(b) [or 2(a) if heating is precisely at the center]. With ``shallow heating'' $P_\mathrm{dep} < P_\mathrm{RCB}$ and regime 2 begins as 2(d). 

The next issue is how regimes evolve as the planet cools. In considering this issue, we will explain how regime 2(c) arises. The bottom panels of \Fig{fig:cartoon} illustrate possible evolutionary pathways in regime 2, assuming that the heat input remains steady at a fixed depth. In regime 2(d), planetary cooling pushes $P_\mathrm{RCB}$ to greater depths. The planet remains in regime 2(d), as depicted in the bottom left panel. Note that continued cooling, and the associated global radius decrease, will not cause a planet to leave regime 2(d). To see this, note that the temperature-pressure profile is given by $d\mathrm{ln}(T)/d\mathrm{ln}(P) = \mathrm{min}(\nabla_\mathrm{ad},\nabla_\mathrm{rad})$, and that both the adiabatic and radiative gradients are explicitly independent of radius (see Equation \ref{eq:radgrad}). In principle a fully radiative end state to regime 2(d) is possible, but in practice the timescale to reach this state is extremely long. 

Evolution in regime 2(a) is depicted in the bottom central panel of \Fig{fig:cartoon}. The planet never leaves regime 2(a) if the central heating remains steady. Cooling and entropy loss can proceed until the planet reaches steady state with $L_\mathrm{cool} = 0$, as described above. In practice, however, the cooling time simply becomes greater than the age of the system. Following \cite{Arras:2006kl}, we can use the entropy equation to estimate the time-derivative of the characteristic internal entropy, assuming heating in the internal convective region:
\begin{equation}
\frac{dS_c}{dt} \sim \frac{(\Gamma - L)}{T_c M} \mathrm{.}
\end{equation}
The internal entropy $S_c$ always decreases with time, as $L = \Gamma + L_{\mathrm{cool}}  > \Gamma$. However, if $\Gamma \gg L_{\mathrm{cool}}$, the entropy decrease will be effectively zero. This leads to a steady state where $L_{\mathrm{cool}} \rightarrow 0$ and all of the outgoing luminosity from the RCB is supplied by the deposited heat.

Cooling evolution in regime 2(b) is the most complex, as illustrated in the bottom right panel of \Fig{fig:cartoon}. Cooling in regime 2(b) leads to an increase in $P_\mathrm{RCB}$ and corresponding decrease in $L_\mathrm{cool}$. However, $P_\mathrm{RCB}$ cannot be deeper than $P_\mathrm{dep}$ in regime 2(b), by definition, and $L_\mathrm{cool}$ cannot drop below the limit described above, i.e. $L_\mathrm{rad,ad}(P_\mathrm{dep})$. Furthermore, regime 2(b) should not simply evolve into regime 2(d) because with $\Gamma \gg L_\mathrm{cool}$, the heating is sufficient to trigger convection as the cooling drops. Instead, continued cooling in regime 2(b) leads to the opening of a radiative window below $P_\mathrm{dep}$, i.e. a transition to regime 2(c) as depicted in \Fig{fig:cartoon}. Once in regime 2(c), cooling proceeds qualitatively as in 2(d), except with an outer convective layer. Hence, the inner RCB retreats with time and the planet remains in regime 2(c).

In summary, in regime 2 with steady heating a planet that enters regime 2(a), (c), or (d) should remain in that specific regime, while regime 2(b) can transition to regime 2(c). This basic understanding is reflected in our numerical evolutionary models. Clearly more complicated behavior is possible if the heating is very broadly distributed or evolving in time.  These more complicated cases are not our focus, but they can probably be understood with a combination of the regimes described here.  The notable exception is re-inflation, which we do not attempt to describe here because the processes by which a planetary interior gain entropy remain uncertain. Next, in \Sec{sec:results} we describe our numerical results, using the discussion here as a backbone for understanding the various structural regimes present and the evolution between them. 
\section{Results}
\label{sec:results}
\begin{figure}
	\centering
	\includegraphics[width=0.5\textwidth]{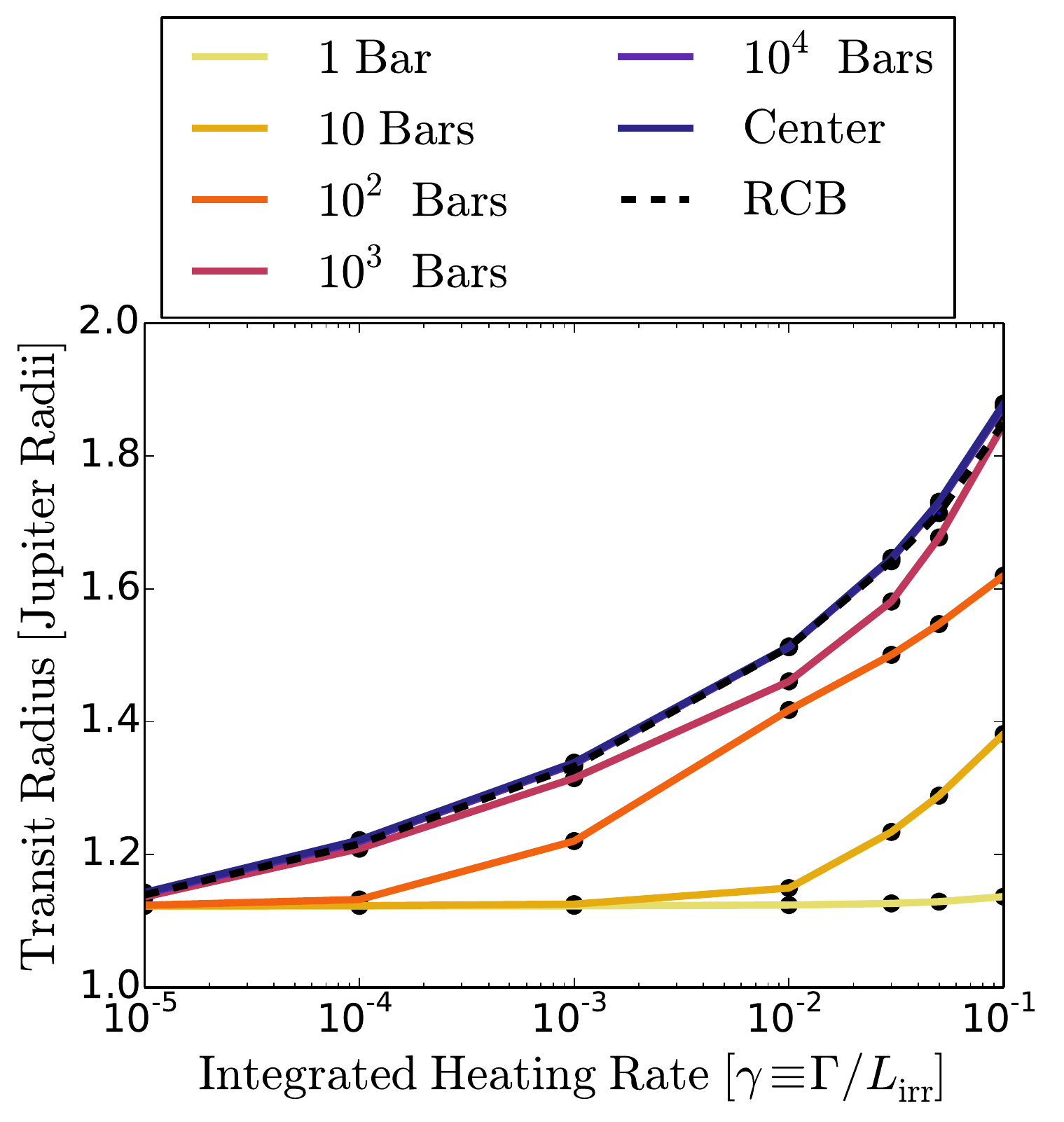}
	\caption{Transit radius in Jupiter radii as a function of normalized integrated heating rate $\gamma \equiv \Gamma/L_{\mathrm{irr}}$ and heating location $P_{\mathrm{dep}}$. Lines correspond to different depths of deposition, colored by pressure, with darker lines corresponding to deeper heating. The line for heating at the radiative-convective boundary is dashed. Scatter points represent the results of individual model runs in our grid. All values are taken at the end of an individual model run, after $5 \ \mathrm{Gyr}$ of evolution. There are two key notable features: similar radii for all cases with heating at $P_{\mathrm{dep}} > 10^3 \ \mathrm{bars}$, and a large ($\sim 25\%$) jump in radius between cases with $P_{\mathrm{dep}} = 10$ and $100 \ \mathrm{bars}$ for $\gamma \ge 10^{-2}$.}
	\label{fig:radcompare}
\end{figure}
\subsection{Radii of Internally Heated Hot Jupiters}
\label{sec:radii}
\Fig{fig:radcompare} shows the transit radius\footnote{To calculate the transit radius from the photospheric radius, we use the isothermal limit of \cite{Guillot:2010} (their Equation 60), setting the ratio of visible to infrared opacities $\kappa_v/\kappa_{th} = 0.4$.}
after 5 Gyr of evolution for our grid of simulations varying $P_{\mathrm{dep}}$ and $\Gamma$. Generally, we find that the final radius increases monotonically with increasing $P_{\mathrm{dep}}$ and $\Gamma$, as expected from previous numerical \citep{Spiegel:2013} and analytic \citep{Ginzburg:2015} work. However, there two key interesting features imprinted on top of this trend. \\
\begin{figure}
	\centering
	\includegraphics[width=0.5\textwidth]{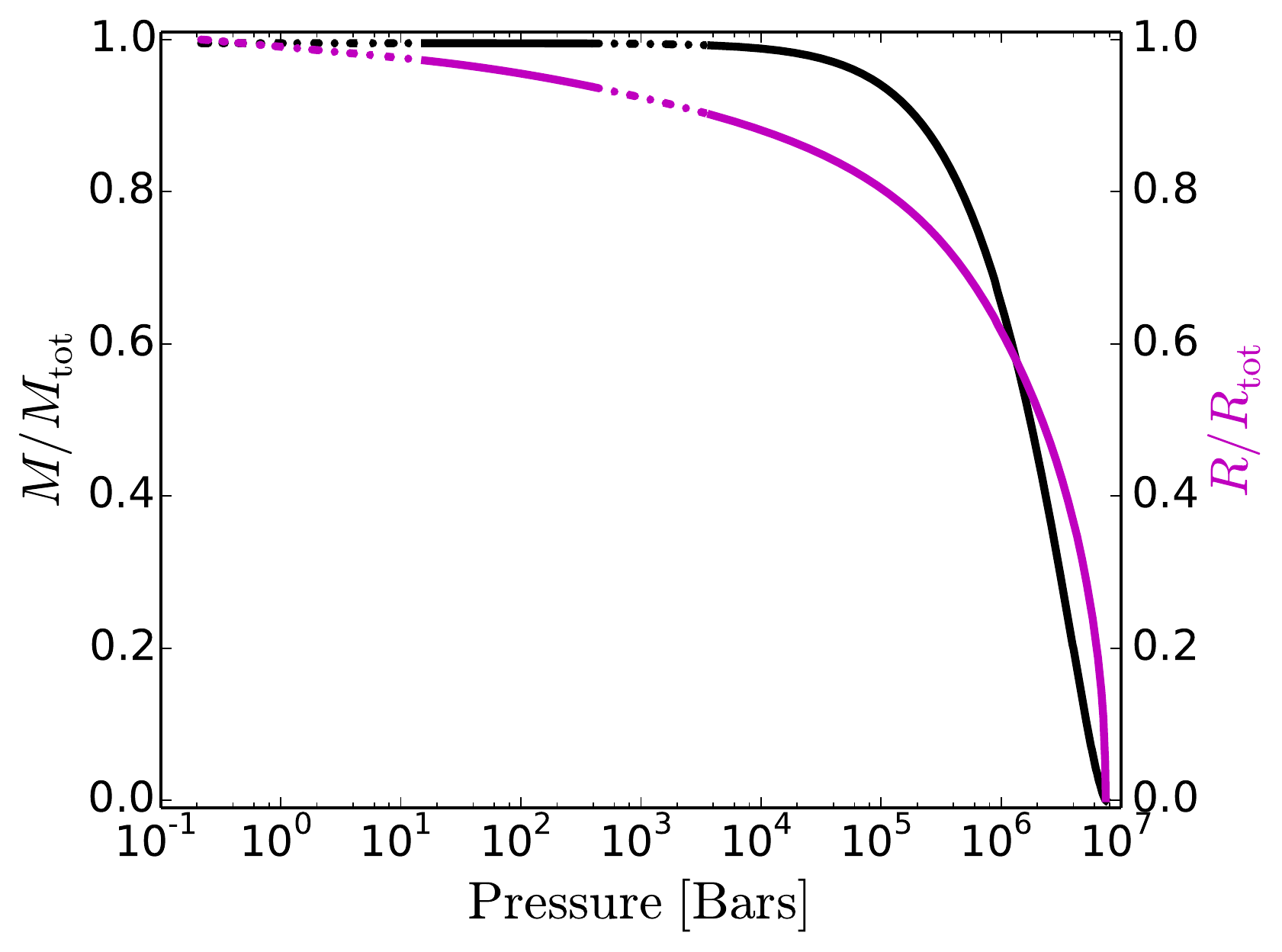}
	\caption{Mass (black) and radius (magenta) vs. pressure profiles at 5 Gyr for the model with normalized heating rate $\gamma = 10^{-2}$ and $P_{\mathrm{dep}} = 100 \ \mathrm{bars}$. The mass and radius are normalized to that of the outermost grid cell. Solid lines show convective zones, and dashed-dotted lines show radiative zones. The heating level ($100 \ \mathrm{bars}$) contains $> 99.99\%$ of the mass and $>95\%$ of the radius of the planet. }
	\label{mass_rad}
\end{figure}
First, there is a large increase in transit radius between the cases with heating at $P_{\mathrm{dep}} = 10$ and  $100 \ \mathrm{bars}$. Secondly, the model transit radii are essentially the same for all of the cases with deeper heating, namely with heating either at $P_{\mathrm{dep}} = 10^4 \ \mathrm{bars}$, heating at the center, or heating at the RCB. 
\Fig{mass_rad} shows that heating at $100 \ \mathrm{bars}$ still encloses $> 99.99\%$ of the mass of the planet. Even heating at $10^4 \ \mathrm{bars}$, which is in the inner convective region, encloses $\approx 99\%$ of the mass. Hence, heating at very shallow regions can greatly affect evolution, in some cases having a similar effect to heating at the center of the planet. We will focus on these two key features in most of the discussion on internal structure and evolution that follows.
\subsection{Structure \& Evolution}
\label{sec:structevolve}
\subsubsection{Structure: Varying Depth of Heating}
\label{sec:structonly}
\begin{figure}
	\centering
	\includegraphics[width=0.5\textwidth]{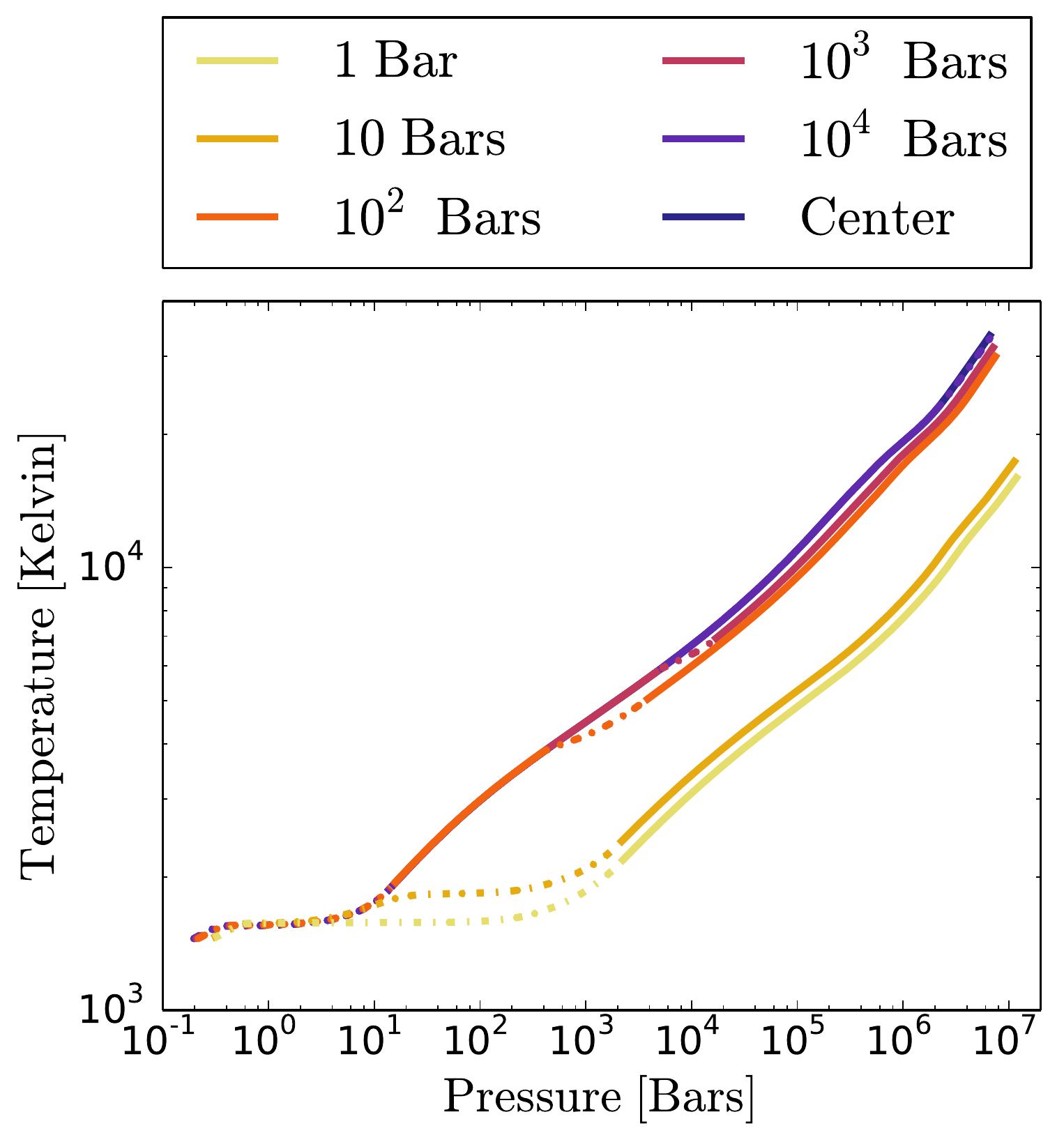}
	\caption{Temperature-pressure profiles at 5 Gyr for slice of runs with normalized heating rate $\gamma = 10^{-2}$ and varying $P_{\mathrm{dep}}$. Dash-dotted regions are radiative zones, and solid regions convective zones. There is a transition between models without an external convective zone with $P_{\mathrm{dep}} \le 10$ bars and with a detached convective zone at $P_{\mathrm{dep}} \ge 100$ bars, leading to a much larger entropy of the internal adiabat in the $P_{\mathrm{dep}} = 100$ bars case relative to that with $P_{\mathrm{dep}} = 10$ bars.}
	\label{fig:TP_1p}
\end{figure}
\begin{figure*}
	\centering
	\includegraphics[width=1\textwidth]{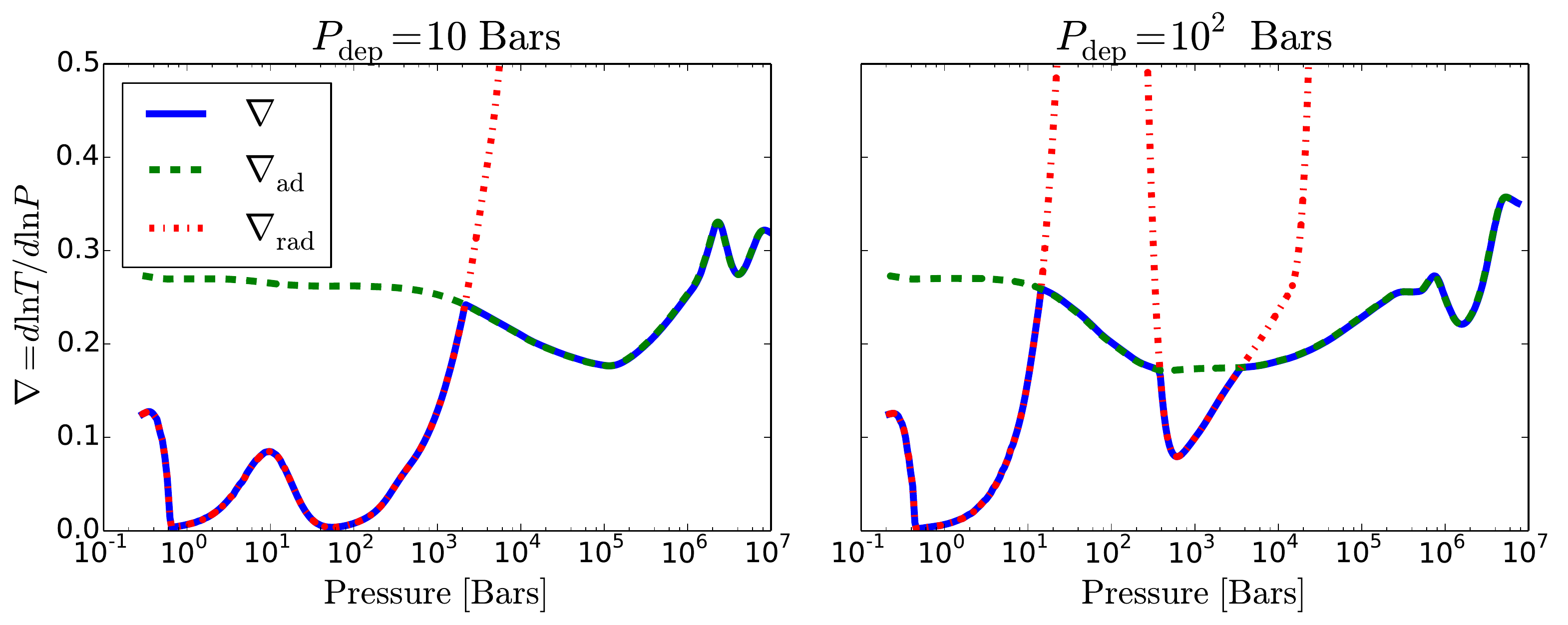}
	\caption{Profiles of actual (blue solid line), adiabatic (green dashed line), and radiative (red dot-dashed line) gradients $\nabla = d\mathrm{ln}T/d\mathrm{ln}P$ as a function of pressure for simulations with $\gamma = 10^{-2}$ and $P_{\mathrm{dep}} = 10, 100 \ \mathrm{bars}$ after 5 Gyr of evolution. The case with $P_{\mathrm{dep}} = 10$ bars has a large external radiative zone, while the model with $P_{\mathrm{dep}} = 100$ bars has a detached outer convective zone due to the large radiative gradient in the heating region.}
	\label{fig:nabla_1p}
\end{figure*}
The trends in transit radius with varying $P_{\mathrm{dep}}$ and $\gamma$ shown in \Fig{fig:radcompare} can be understood by examining the internal structure. Our goal is to understand the two specific features introduced above. First, all cases with $P_{\mathrm{dep}} \gtrsim 10^4 \ \mathrm{bars}$ show almost the same radius inflation for a given heating rate $\gamma$. Secondly, there is a large ($\sim 25\%$) jump in radius between $P_{\mathrm{dep}} = 10 - 100 \ \mathrm{bars}$. As we show below, the larger radius with deeper heating corresponds to the opening of an internal radiative window, i.e. cooling in regime 2(c) instead of regime 2(d), as introduced in \Sec{sec:cooling}. \\
\indent \Fig{fig:TP_1p} shows pressure-temperature profiles at the end-state of evolution for varying $P_{\mathrm{dep}}$ with a fixed normalized heating rate $\gamma = 10^{-2}$. These profiles show a clear bifurcation between shallow heating ($P \le 10 \ \mathrm{bars}$) and deep heating ($P \ge 100 \ \mathrm{bars}$). Simulations with shallow heating at $P_{\mathrm{dep}} = 1 - 10 \ \mathrm{bars}$ have a deep outer radiative zone, extending to $P_\mathrm{RCB} > 10^3 \ \mathrm{bars}$. By contrast, with deep heating at $P_{\mathrm{dep}} \ge 100 \ \mathrm{bars}$ the outer radiative zone only extends to $P \sim 10 \ \mathrm{bars}$. The shallow cases have cooled in regime 2(d) while the deep cases end in regime 2(c), or 2(a) with heating directly at the center. With deeper heating, the transition to a steeper adiabatic profile at lower pressures clearly allows the planet to reach a higher central temperature and thus central entropy and transit radius. Ultimately though, it is the suppression of cooling by the deeper heating that allows this higher internal entropy at late times. \\
\indent Note that the outer convective zones with $100 \ \mathrm{bars} \le P_{\mathrm{dep}} \le 10^4 \ \mathrm{bars}$ do not match onto the same internal adiabat. Deep heating forces outer convective zones with corresponding internal radiative windows, as found in both numerical \citep{Guillot_2002,Batygin_2011,Wu:2013} and analytic \citep{Ginzburg:2015} studies. Radiative windows require the internal adiabat to have lower entropy than the outer convective zone \citep{Guillot:1994a}. Nevertheless, the central entropy still remains much larger than for the shallow heating cases which fail to trigger outer convective zones. \\
\indent  \Fig{fig:nabla_1p} shows profiles of $\nabla,\nabla_{\mathrm{ad}},\nabla_{\mathrm{rad}}$ for simulations with the normalized heating rate $\gamma = 10^{-2}$. The two cases of $P_\mathrm{dep} = 10$ and $100 \ \mathrm{bars}$ illustrate the transition from shallow to deep heating, respectively. The profiles in \Fig{fig:nabla_1p} show that heating at $P_{\mathrm{dep}} = 100 \ \mathrm{bars}$ triggers an outer convective region, with an inner radiative window below. Since $\nabla_{\mathrm{rad}} \propto LP$, the same heating ($L$) at a deeper pressure forces a secondary convective region.
For heating at $P_{\mathrm{dep}} = 10 \ \mathrm{bars}$, the increase in $\nabla_{\mathrm{rad}}$ is too small to trigger convective instability. \\
\indent The amount of heat required to force a detached outer convective zone can be quantified as \citep{Ginzburg:2015}:
\begin{equation}
\Gamma \tau_{\mathrm{dep}} \sim \Gamma \frac{P_{\mathrm{dep}} \kappa}{g} \gtrsim L_{\mathrm{irr}} \mathrm{.}
\end{equation}
The quantity $\Gamma \tau_{\mathrm{dep}} \sim 10^2 L_{\mathrm{irr}}$ for $P_{\mathrm{dep}} = 100 \ \mathrm{bars}$ and $\Gamma \tau_{\mathrm{dep}} \sim L_{\mathrm{irr}}$ for $P_{\mathrm{dep}} = 10 \ \mathrm{bars}$, which is hence just shallow enough to not force a secondary convective zone. As a result, the simulations with $P_{\mathrm{dep}} \le 10 \ \mathrm{bars}$ match onto a lower internal adiabat, naturally explaining the $\sim 25\%$ smaller transit radius.
\subsubsection{Evolution: Structural Quantities}
\label{sec:evolve}
\begin{figure*}
	\centering
	\includegraphics[height=1\textheight]{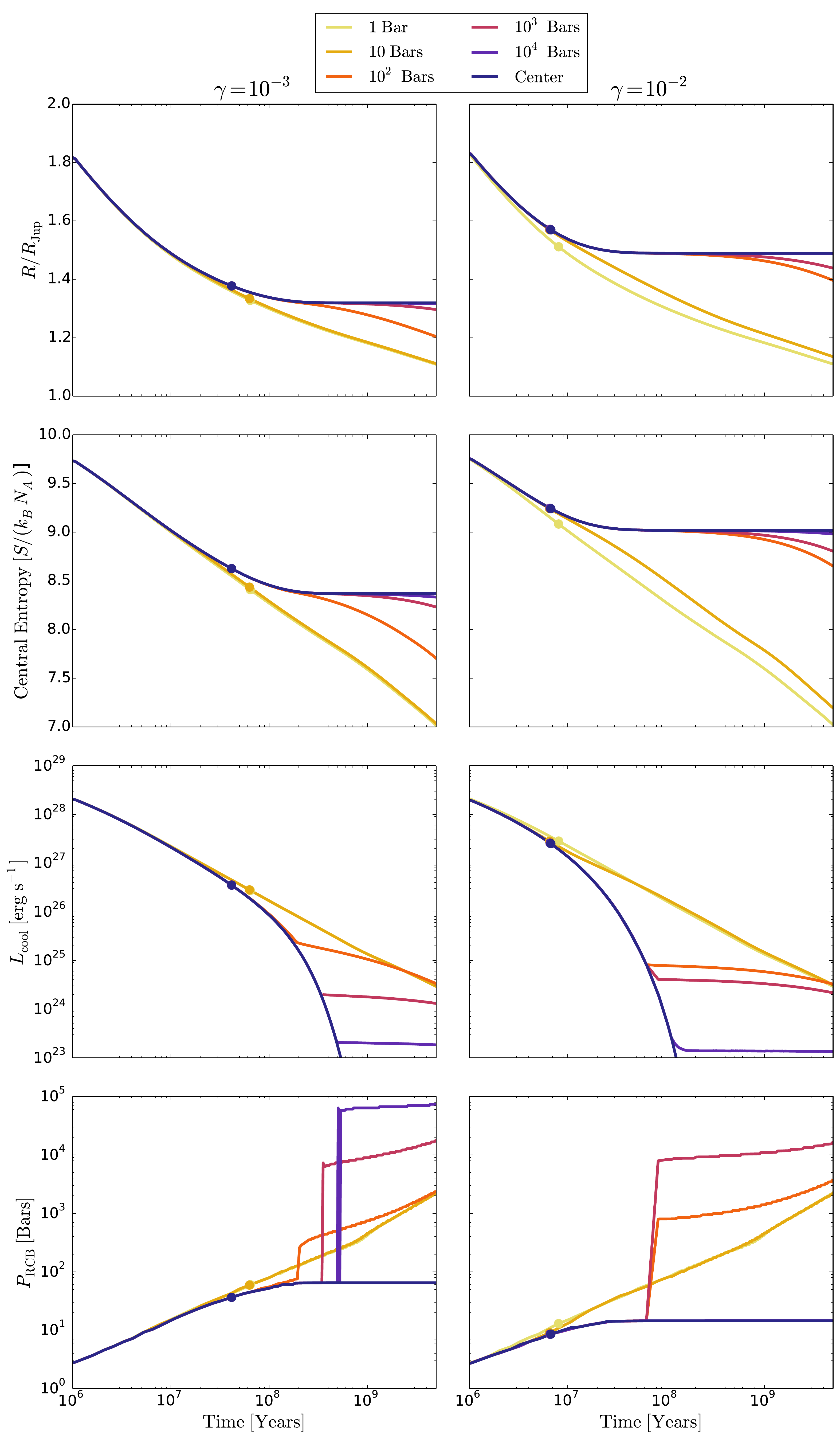}
	\caption{Time evolution of photospheric radius, central entropy, total internal cooling luminosity, and inner radiative-convective boundary pressure for simulations with normalized heating rates $\gamma = 10^{-3},10^{-2}$ and varying $P_{\mathrm{dep}}$. The colored dots on each line mark the transition between the regimes where the internal cooling luminosity $L_{\mathrm{cool}}$ is greater than (at times before the dot) or less than (times after the dot) the integrated heating level $\Gamma$.
	Note the correspondence between the time at which the cooling luminosity (equivalent to $L_{\mathrm{cool}}$) flattens out and the appearance of an inner radiative window.}
	\label{fig:compare_time}
\end{figure*}
Thus far we have only presented the end-state of evolution, without examining any temporal changes in structural quantities. However, examining the cooling history of our models is crucial for understanding trends with varying $P_{\mathrm{dep}}$ and $\Gamma$. Here we investigate the evolution of two slices of our grid, varying $P_{\mathrm{dep}}$ with fixed normalized heating rate $\gamma = 10^{-2},10^{-3}$. \Fig{fig:compare_time} displays the time-evolution of photospheric radius, central entropy, the gravitational cooling luminosity
\begin{equation}
 L_{\mathrm{cool}} = \int_0^M -T \frac{dS}{dt} dm \mathrm{,}
 \end{equation}
 and the pressure at the inner RCB for $\gamma = 10^{-2}$ and $10^{-3}$. \\
\indent The early evolution of these structural quantities is similar for both $\gamma = 10^{-2}$ and $10^{-3}$. All models show an identical cooling phase over the first $\sim 10 \ \mathrm{Myr}$, which extends to $\sim 50 \ \mathrm{Myr}$ with weaker heating of $\gamma = 10^{-3}$. This uniform cooling phase is called regime 1 in \Sec{sec:cooling}. This phase ends when the gravitational cooling luminosity falls below the applied heating rate, explaining why this cooling phase lasts longer with weaker heating. After this point, the evolution diverges based on whether the heating is shallow (i.e. at $1 - 10 \ \mathrm{bars}$) or deep ($\ge 100 \ \mathrm{bars}$). \\
\indent When heating is shallow, the planet continuously cools over time in regime 2(d) (i.e. with heating in the radiative exterior). The central entropy and thus planetary radius decrease smoothly with time. The RCB gradually retreats deeper into the planet, which causes the smooth decline in cooling rate. \\
\indent For cases with deeper heating, $P_\mathrm{dep} \ge 100 \ \mathrm{bars}$, the planet next reaches a quasi-equilibrium state, during which the planetary radius and entropy remain (nearly) constant over some period of time. The bottom panel of \Fig{fig:compare_time} shows that during this equilibrium the interior RCB lies outside (at lower pressure than) the heating zone. For the case of $P_\mathrm{dep} = 100 \ \mathrm{bars}$, $P_\mathrm{RCB} \sim 15$ and $70 \ \mathrm{bars}$ for $\gamma = 10^{-2}$ and $10^{-3}$ in this state, respectively. Thus the hot Jupiters are mostly in regime 2(b) during this quasi-equilibrium phase. \\
\indent The central heating case, i.e. regime 2(a), stays in the near equilibrium state. This central heating case rapidly approaches a true equilibrium with $L_\mathrm{cool} \rightarrow 0$ in which heating supplies the total interior luminosity. This behavior is evident in the sharp drop-off in $L_\mathrm{cool}$, see the third row of \Fig{fig:compare_time}. \\
\indent For the cases with deep atmospheric heating at $10^2 \ \mathrm{bars} \le P_\mathrm{dep} \le 10^4 \ \mathrm{bars}$, the quasi-equilibrium state eventually comes to an end and the planet begins cooling and contracting again. The bottom row of \Fig{fig:compare_time} shows that the departure from quasi-equilibrium roughly corresponds to a sharp increase in $P_\mathrm{RCB}$, i.e. the opening of a radiative window which marks the transition from regime 2(b) to 2(c). When the radiative window opens, $L_\mathrm{cool}$ does not increase, but it stops declining rapidly with time. This more slowly evolving $L_\mathrm{cool}$ is sufficient to eventually cause a noticeable decrease in planetary entropy and radius. \\
\indent The case with $P_\mathrm{dep} = 10^4 \ \mathrm{bars}$ is notable because the evolution in radius and central entropy is nearly indistinguishable from the case with central heating. The behavior of $L_\mathrm{cool}$ is noticeably different for the two heating locations, with $L_\mathrm{cool}$ dropping to arbitrarily low levels at late times with applied central heating. However, with $P_\mathrm{dep} = 10^4 \ \mathrm{bars}$, $L_\mathrm{cool}$ stays just above $10^{23} \mathrm{erg} \ \mathrm{s}^{-1}$, for both $\gamma = 10^{-2}$ and $10^{-3}$. Crucially, this finite level of cooling corresponds to Kelvin-Helmholtz cooling times of $\sim 10^3 \ \mathrm{Gyr}$, much longer than the system age. Thus, departures from a true equilibrium with no cooling are hard to discern. \\
\indent A subtle point about the $P_\mathrm{dep} = 10^4 \ \mathrm{bars}$ case is that the floor in $L_\mathrm{cool}$ at late times has a different origin in the $\gamma = 10^{-3}$ and $10^{-2}$ cases, despite their similar value of $L_\mathrm{cool}$. For $\gamma = 10^{-3}$, the plateau in $L_\mathrm{cool}$ occurs when a radiative window opens. This window opens at $\sim 500 \ \mathrm{Myr}$, i.e. $\sim 10\%$ of the final age. However, the window is so deep, approaching a pressure of $10^5 \ \mathrm{bars}$, that cooling is very slow. This example emphasizes that the opening of the radiative window [i.e. entering regime 2(c)] does not necessarily correspond to measurable radius contraction over stellar lifetimes. For $\gamma = 10^{-2}$, a radiative window does not open. In this case the floor in $L_\mathrm{cool}$ corresponds to the minimum value of $L_\mathrm{rad,ad}(P_\mathrm{dep})$ discussed in \Sec{sec:cooling}. 
\subsubsection{Evolution: Luminosity and Temperature Profiles}
\label{sec:evolprofiles}
\begin{figure}
	\centering
	\includegraphics[width=0.5\textwidth]{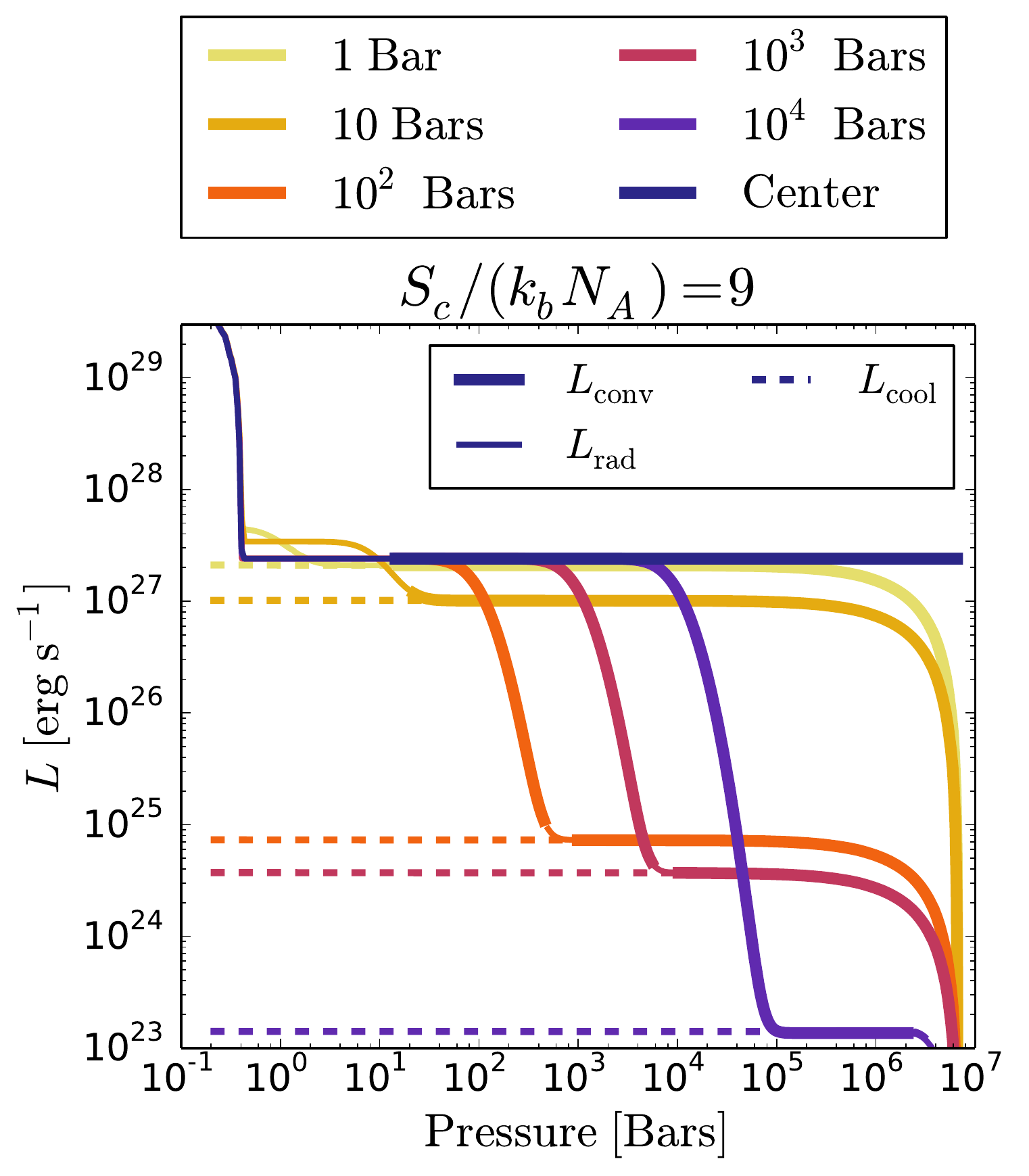}
	\caption{Luminosity-pressure profiles at fixed central entropy $S_c/(k_bN_A) = 9$ for a normalized heating rate $\gamma = 10^{-2}$ and varying $P_{\mathrm{dep}}$. Thick solid lines show the luminosity in convective zones, and thin sold lines that in radiative zones. The dashed lines show the luminosity from internal cooling alone, i.e. $L_{\mathrm{cool}}$.
	There are three main features, from the center of the planet outward: a rise to the internal cooling rate $L_{\mathrm{cool}}$, a second rise to the heating level $\Gamma$, and a third rise to the incident stellar power $L_{\mathrm{irr}}$. In the case with heating at the very center, $\Gamma \gg L_{\mathrm{cool}}$ and only the second rise to $\Gamma$ is seen.}
	\label{fig:1p_Seq9}
\end{figure}
\begin{figure*}
	\centering
	\includegraphics[width=0.7\textwidth]{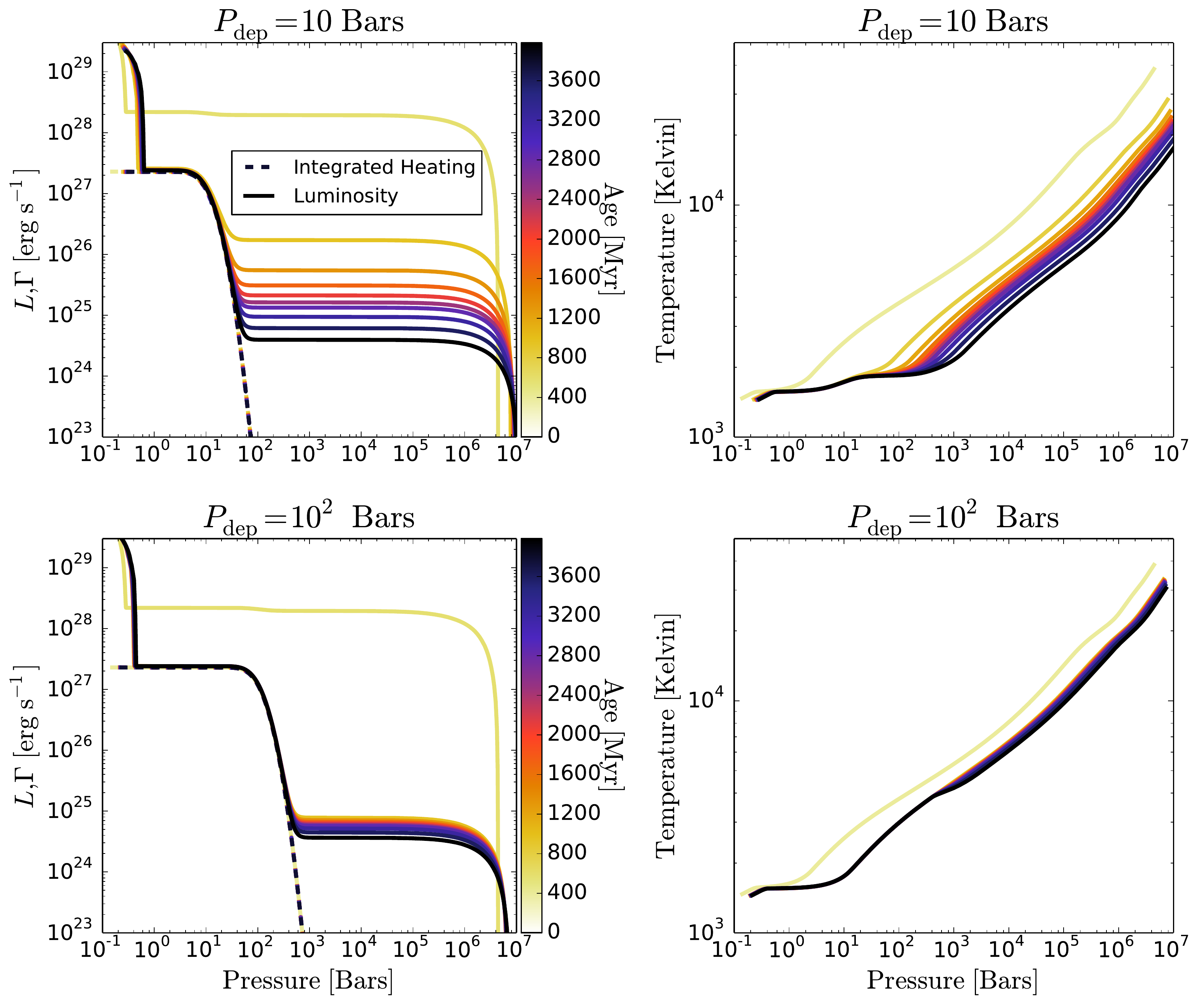}
	\caption{Time-evolution of luminosity $L$ \& integrated heating $\Gamma$ (left) and temperature (right) profiles for simulations with a normalized heating rate $\gamma = 10^{-2}$ and $P_{\mathrm{dep}} = 10 \ \mathrm{bars}$ (top) and $100 \ \mathrm{bars}$ (bottom). The case with $P_{\mathrm{dep}} = 10 \ \mathrm{bars}$ cools from a deep outer radiative zone, while the case with $P_\mathrm{dep} = 100 \ \mathrm{bars}$ cools from an inner RCB at the base of a radiative window. As a result, the inner adiabat with $P_\mathrm{dep} = 100 \ \mathrm{bars}$ has greatly reduced cooling relative to the case with shallower heating.}
	\label{fig:L_T_P_1p}
\end{figure*}
\indent To understand in detail how the pressure at which heat is deposited changes the internal cooling rate, we examine runs at fixed central entropy. Though these model planets have the same central entropy and radius, we are sampling different times in their evolution and hence different internal cooling rates. \Fig{fig:1p_Seq9} shows luminosity-pressure profiles for a normalized heating rate $\gamma= 10^{-2}$ and varying deposition pressure at fixed central entropy $S_c/(k_bN_A) = 9$, which is approximately the equilibrium entropy for central heating at this value of $\gamma$. Luminosity is an integrated quantity from the center, and as a result it monotonically increases outward. There are three main features in this luminosity profile, from the center outwards: an initial rise to the level of $L_{\mathrm{cool}}$, a secondary rise to the heating level $\Gamma$, and a third to the irradiation level $L_{\mathrm{irr}}$. The profiles in the outer two levels are almost the same for all $P_{\mathrm{dep}}$, but the internal cooling rates (shown by the dashed lines) vary by four orders of magnitude across the sample. This is because if heating  is deep enough to affect the internal structure, radiative losses effectively occur from just below the heating layer. The cooling rate then decreases with increasing $P_\mathrm{dep}$ (at fixed central entropy), due to the longer cooling timescales at greater depths. \\
\indent To examine further the differences between the cases with $P_{\mathrm{dep}} = 10$ and $100 \ \mathrm{bars}$, we show the time-evolution of their luminosity and temperature profiles with $\gamma = 10^{-2}$ in \Fig{fig:L_T_P_1p}.  During the free cooling phase, which corresponds to regime 1, the cooling luminosity is much larger than the integrated heating rate and the heating does not affect the temperature structure. In regime 2, the planet has cooled such that the integrated heating rate is larger than the internal cooling rate. As a result, the luminosity and temperature profiles are essentially fixed in time at levels shallower than the deposition pressure $P_{\mathrm{dep}}$. In this regime, the cooling itself mostly occurs at the RCB, which corresponds to the time-evolution of $L_{\mathrm{cool}}$ shown in \Fig{fig:compare_time}. This cooling forces the RCB to move inward with time. In the case with deeper $P_\mathrm{dep} = 100 \ \mathrm{bars}$ there is an outer convective zone and hence this cooling occurs at the inner RCB. Just above this inner RCB, there is a deep near-isothermal region, which corresponds to the location of a radiative window. This radiative window occurs just below the heating level, and the temperature at the top of the radiative window is determined by the integrated heating rate. Hence, even though the integrated cooling rates after $5 \ \mathrm{Gyr}$ for the cases with $P_{\mathrm{dep}} = 10$ and $100 \ \mathrm{bars}$ are similar, the radius for $P_{\mathrm{dep}} = 100 \ \mathrm{bars}$ is larger. The larger radius is due to the much higher temperature at the bottom of the innermost radiative zone, which leads to a hotter internal adiabat. 
\section{Comparison with Analytic Theory}
\label{sec:discussion}
\begin{figure}
	\centering
	\includegraphics[width=0.5\textwidth]{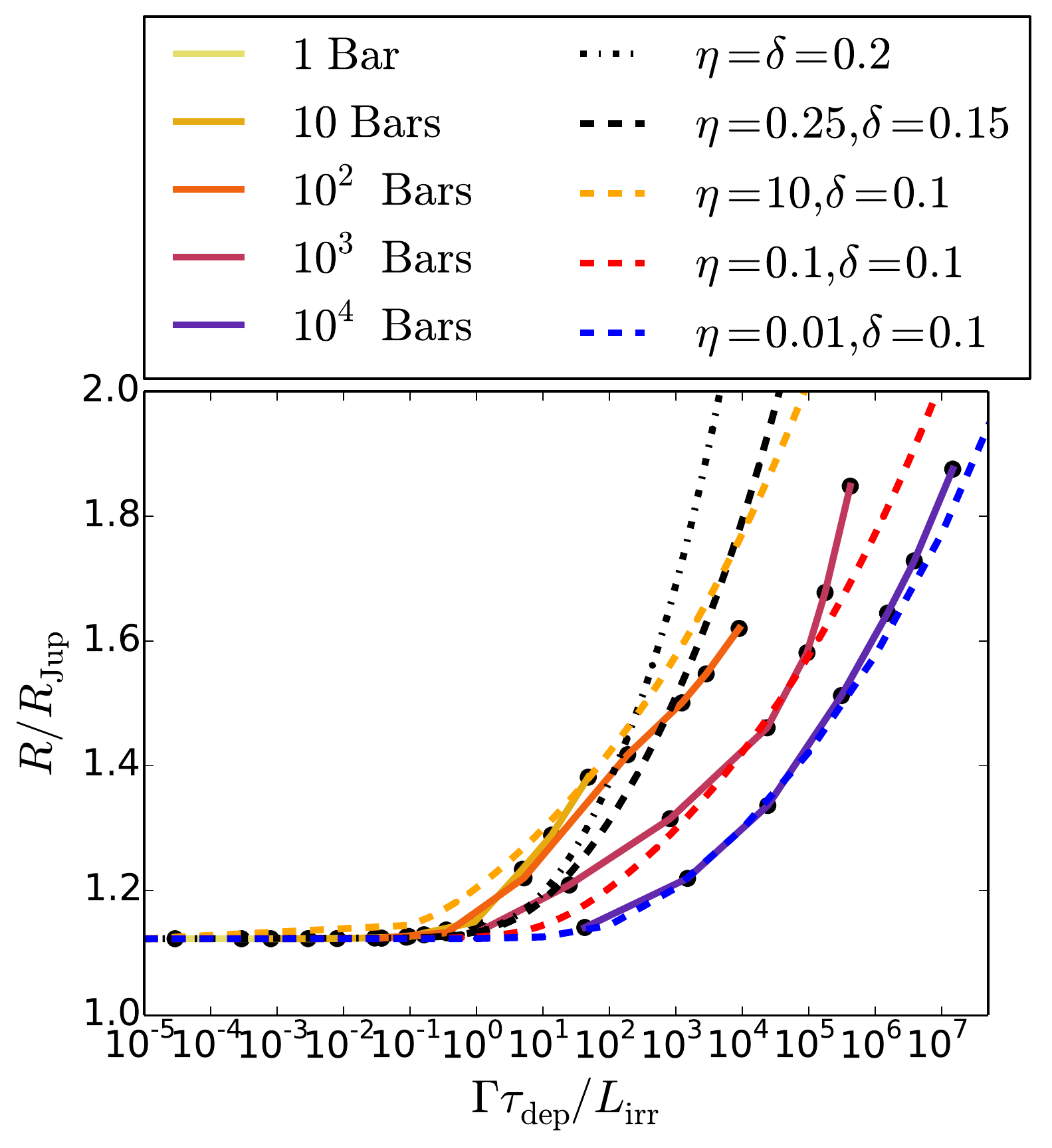}
	\caption{Transit radius in Jupiter radii at 5 Gyr as a function of $\Gamma \tau_{\mathrm{dep}}$, normalized by the incident stellar power. Solid lines show numerical results for various $P_{\mathrm{dep}}$, while dashed and dashed-dotted lines show analytic predictions from \Eq{eq:rpredict}. The two black lines show results for the $\eta, \delta$ values used in \cite{Ginzburg:2015}. The dashed colored lines show illustrations of the fit to the R-$\Gamma \tau_{\mathrm{dep}}$ relationship, which requires the same choice of $\delta$ but decreasing $\eta$ with increasing $P_{\mathrm{dep}}$.}
	\label{fig:Ltaudep}
\end{figure}
\indent In this section, we compare our numerical models to the analytic theory of \cite{Ginzburg:2015}. 
They related the quantity $\Gamma \tau_{\mathrm{dep}}$ to the radius of a hot Jupiter given specified external parameters (equilibrium temperature, planet mass, composition), as it is linked to the central temperature. Specifically, their prediction for radius given $\Gamma \tau_{\mathrm{dep}}$ (equivalent to Equation (32) of \citealp{Ginzburg:2015}) is 
\begin{equation}
\label{eq:rpredict}
R = R_0 + \Delta R_0 \left[\left(1 + \eta \frac{\Gamma \tau_{\mathrm{dep}}}{L_{\mathrm{irr}}}\right)^{\delta} -1 \right] \mathrm{,}
\end{equation}
where $R_0$ is the radius without extra heat deposition ($\approx 1.1 \ R_{\mathrm{Jup}}$), $\Delta R_0 = 0.3 R_{\mathrm{Jup}}$, $\tau_{\mathrm{dep}}$ the optical depth at which heat is deposited, and $\eta$ and $\delta$ fitting parameters. Note that this prediction is for the radius of a planet with a given internal luminosity, while our numerical simulations calculate the radius at a given age. \cite{Ginzburg:2015} make no explicit prediction for radius at a given age, instead predicting radius for a given cooling luminosity to compare with the numerical calculations of \cite{Spiegel:2013}. Despite the fact that \cite{Ginzburg:2015} make no explicit prediction for the radius as a function of time, here we test the utility of their scaling $\Delta R \propto T_\mathrm{c} \propto \left(1+ \Gamma \tau_{\mathrm{dep}}/L_{\mathrm{irr}}\right)^{\delta}$ as motivated by their Equations (23) and (29). \\
\indent The power-law exponent $\delta$ can be determined analytically from the use of a power-law opacity and relationship between radiation energy density and optical depth (see Equation 31 of \citealp{Ginzburg:2015}), but here we treat it as a free parameter due to the use of full opacity tables. \cite{Ginzburg:2015} predicted that $\delta \approx 0.19$ to calculate radius at a given age (see their Equation 23), while $\delta \approx 0.15$ to calculate radius at a given internal luminosity. \\
\indent \Fig{fig:Ltaudep} compares \Eq{eq:rpredict} for various $\eta$ and $\delta$ to our numerically determined radius-$\Gamma \tau_{\mathrm{dep}}$ relationship. First, note that as expected from \Eq{eq:rpredict}, the radius starts to deviate sharply from $R_0$ in our numerical solutions when $\Gamma \tau_{\mathrm{dep}}/L_{\mathrm{irr}} \ge 1$. When $\Gamma \tau_{\mathrm{dep}} < L_{\mathrm{irr}}$, the heat source does not strongly perturb the radiative-convective solution due to the relative dominance of the external irradiation. With $\Gamma \tau_{\mathrm{dep}} \ge L_\mathrm{irr}$, one can think of the heat source as an effective increase of the incident stellar power, which then perturbs the radiative-convective solution. \\
\indent In the regime with $\Gamma \tau_{\mathrm{dep}} \ge L_\mathrm{irr}$, the analytic prediction from \Eq{eq:rpredict} is that the effect of heating increases with increasing $\tau^\delta_\mathrm{dep}$ and hence approximately increases as $P^\delta_\mathrm{dep}$. In the case with $P_\mathrm{dep} \leq 100 \ \mathrm{bars}$, we find that as predicted by \cite{Ginzburg:2015} there is a nearly universal relation between radius and $\Gamma \tau_{\mathrm{dep}}$. We also find that our numerical radius-$\Gamma \tau_{\mathrm{dep}}$ relationship can be reproduced using a constant $\delta \approx 0.1$, which is expected given that $\delta$ is prescribed by the opacity-pressure-temperature relationship \citep{Ginzburg:2015}. Note that our effective $\delta$ is somewhat lower than the $\delta \approx 0.19$ predicted by \cite{Ginzburg:2015} for comparison at equal ages. This is because of the use of full \mesa \ opacities rather than prescribing a power-law opacity. \\
\indent However, there is a deviation between our numerical models and the predictions of \cite{Ginzburg:2015} in the case of deep heating at $P_\mathrm{dep} > 100 \ \mathrm{bars}$. In this regime, we need significantly smaller $\eta$ values to match the radius for larger values of $P_\mathrm{dep}$\footnote{Note that the $\eta$ needed becomes tiny ($\sim 10^{-6}$) in the case with heating at the very center of the planet (not shown). This case was not examined in detail by \cite{Ginzburg:2015}, who instead focused on $\tau_{\mathrm{dep}} \lesssim \tau_{\mathrm{RCB}}$. The case with heating at the center of the planet does, however, have the same $\delta$, showing that the opacity-pressure-temperature relationship still determines the power of the relationship between $\Gamma \tau_{\mathrm{dep}}$ and radius even with heating at the very center of the planet.}. Thus, the effects of increased heating depths on radius inflation are modest for $P_\mathrm{dep} > 100 \ \mathrm{bars}$. This finding is consistent with our main conclusions, which further show a complete independence of radius inflation on $P_\mathrm{dep} \ge 10^4 \ \mathrm{bars}$. Note that this deviation between simulations and theory is not necessarily a disagreement, as \Eq{eq:rpredict} is not directly applicable deeper than the radiative-convective boundary. In this case where $\tau_\mathrm{dep} > \tau_\mathrm{RCB}$, it is required that $\Gamma > L_\mathrm{cool}$ in order for heating to significantly affect the radius \citep{Guillot_2002,Wu:2013,Ginzburg:2015}. However, \cite{Ginzburg:2015} did not develop an analytic theory to compare with numerical results with $P_\mathrm{dep} > 100 \ \mathrm{bars}$, because these results did not exist at the time. We thus show that their model is very useful for shallow heating, but should not be used for a quantitative prediction of radius at a given age for heat deposited deeper than $100 \ \mathrm{bars}$. 
\vspace{1.25cm}
\section{Conclusions}
\label{sec:conclusions}
To better understand the transit radii of hot Jupiters, we studied the evolution of a giant planet subject to intense stellar irradiation at the surface and internal heating. To do so, we used \mesa \ to compute a grid of evolutionary models in which we varied the amount and depth of internal heating using the parameters of a typical hot Jupiter, HD 209458b. To interpret our numerical results, we developed a framework to understand the different cooling regimes of internally heated hot Jupiters. Based on previous work, it is known that the radii of hot Jupiters is larger for more intense \textit{or} deeper internal heating, provided that this heating is applied throughout the planets' evolution. Our results are broadly consistent with this expectation, but show that this trend with heating depth, while monotonic, is very uneven. Specifically, we find that:
\begin{enumerate}
\item A minimum heating depth of $100 \ \mathrm{bars}$ is required to explain inflated hot Jupiter radii, assuming modest internal heating rates $\lesssim 1\%$ of the stellar irradiation. This pressure level is deeper than the photosphere yet is overall very shallow, enclosing $99.99\%$ of the mass of the planet. We show that modest heating at $\gtrsim 100 \ \mathrm{bars}$ is deep enough to enhance the radius for two reasons. First, this heating lies within the convective interior for the initial $10-100 \ \mathrm{Myr}$ of evolution. Cooling is thereby significantly suppressed during this crucial early stage. Second, at later times a radiative window opens below the heating layer. The cooling rate of the planet, set at the base of this radiative window, is very low. This enables the planet to retain a much larger radius than for shallower heating. 
\item Heating applied at any depth $\gtrsim 10^4 \ \mathrm{bars}$ yields nearly identical levels of radius inflation. Since $\approx 99\%$ of the mass of a hot Jupiter lies below $10^4 \ \mathrm{bars}$, it is remarkable that deeper heating -- even heating at the center -- produces a similar cooling history. The reason for this depth independence is that cooling timescales from these deep pressures exceed the several Gyrs lifetime of the planet. 
\end{enumerate} 
\indent Since our models are agnostic as to the mechanism of internal heating, they can be used to constrain a range of hot Jupiter heating mechanisms. Most notably, because we find that at minimum modest heating must occur at $\sim100 \ \mathrm{bars}$ to explain radius inflation, even relatively shallow hydrodynamic inflation mechanisms can explain the transit radii of many hot Jupiters.
\acknowledgements
We thank Sivan Ginzburg, Bill Hubbard, Adam Showman, and the Steward Observatory Planet Theory group for insightful discussions. We thank the anonymous referee for a thorough review, which improved the manuscript. We thank the \mesa \ team for making this valuable tool publicly available. This research has made use of the Exoplanet Orbit Database and the Exoplanet Data Explorer at \url{exoplanets.org}. T.D.K. acknowledges support from NASA headquarters under the NASA Earth and Space Science Fellowship Program Grant PLANET14F-0038. A.N.Y. acknowledges support from the NASA ATP program through grant NNX16AB26G.
\if\bibinc n
\bibliography{References}

\begin{thebibliography}{72}
\expandafter\ifx\csname natexlab\endcsname\relax\def\natexlab#1{#1}\fi

\bibitem[{Arras \& Bildsten(2006)}]{Arras:2006kl}
Arras, P. \& Bildsten, L. 2006, The Astrophysical Journal, 650, 394

\bibitem[{Arras \& Socrates(2010)}]{Arras:2010}
Arras, P. \& Socrates, A. 2010, The Astrophysical Journal, 714, 1

\bibitem[{Baraffe {et~al.}(2010)Baraffe, Chabrier, \& Barman}]{Baraffe:2010xe}
Baraffe, I., Chabrier, G., \& Barman, T. 2010, Reports on Progress in Physics,
  76, 30

\bibitem[{Baraffe {et~al.}(2014)Baraffe, Chabrier, Fortney, \&
  Sotin}]{Baraffe:2014}
Baraffe, I., Chabrier, G., Fortney, J., \& Sotin, C. Protostars and Planets VI,
  ed. H.~Beuther, R.~Klessen, C.~Dullemond, \& T.~Henning (Tucson, AZ:
  University of Arizona Press)

\bibitem[{Batygin \& Stevenson(2010)}]{Batygin_2010}
Batygin, K. \& Stevenson, D. 2010, The Astrophysical Journal Letters, 714, L238

\bibitem[{Batygin {et~al.}(2011)Batygin, Stevenson, \&
  Bodenheimer}]{Batygin_2011}
Batygin, K., Stevenson, D., \& Bodenheimer, P. 2011, The Astrophysical Journal,
  738, 1

\bibitem[{Bodenheimer {et~al.}(2007)Bodenheimer, Laughlin, Rozyczka, \&
  Yorke}]{Bodenheimer:2007}
Bodenheimer, P., Laughlin, G., Rozyczka, M., \& Yorke, H. 2007, Numerical
  Methods in Astrophysics (Boca Raton, FL: Taylor \& Francis)

\bibitem[{Bodenheimer {et~al.}(2001)Bodenheimer, Lin, \&
  Mardling}]{Bodenheimer:2001}
Bodenheimer, P., Lin, D., \& Mardling, R. 2001, The Astrophysical Journal, 548,
  466

\bibitem[{Burrows {et~al.}(2007)Burrows, Hubeny, Budaj, \&
  Hubbard}]{Burrows:2007bs}
Burrows, A., Hubeny, I., Budaj, J., \& Hubbard, W. 2007, The Astrophysical
  Journal, 661, 502

\bibitem[{Chabrier \& Baraffe(2007)}]{Chabrier:2007}
Chabrier, G. \& Baraffe, I. 2007, The Astrophysical Journal Letters, 661, 81

\bibitem[{Chandrasekhar(1939)}]{Chandrasekhar:1939}
Chandrasekhar, S. 1939, An Introduction to Stellar Structure (Chicago, IL:
  University of Chicago Press)

\bibitem[{Charbonneau {et~al.}(2000)Charbonneau, Brown, Latham, \&
  Mayor}]{Charbonneau_2000}
Charbonneau, D., Brown, T., Latham, D., \& Mayor, M. 2000, The Astrophysical
  Journal, 529, L45

\bibitem[{Cooper \& Showman(2005)}]{Cooper:2005}
Cooper, C. \& Showman, A. 2005, The Astrophysical Journal Letters, 629, L45

\bibitem[{Demory \& Seager(2011)}]{Demory:2011}
Demory, B. \& Seager, S. 2011, The Astrophysical Journal Supplement Series,
  197, 12

\bibitem[{Fortney {et~al.}(2010)Fortney, Baraffe, \& Militzer}]{fortney_2009}
Fortney, J., Baraffe, I., \& Militzer, B. Exoplanets, ed. S.~Seager (Tucson,
  AZ: University of Arizona Press)

\bibitem[{Fortney {et~al.}(2008)Fortney, Lodders, Marley, \&
  Freedman}]{Fortney:2008}
Fortney, J., Lodders, K., Marley, M., \& Freedman, R. 2008, The Astrophysical
  Journal, 678, 1419

\bibitem[{Fortney {et~al.}(2007)Fortney, Marley, \& Barnes}]{Fortney:2007ta}
Fortney, J., Marley, M., \& Barnes, J. 2007, The Astrophysical Journal, 659,
  1661

\bibitem[{Freedman {et~al.}(2008)Freedman, Marley, \& Lodders}]{Freedman:2008}
Freedman, R., Marley, M., \& Lodders, K. 2008, The Astrophysical Journal
  Supplement Series, 174, 504

\bibitem[{Ginzburg \& Sari(2015)}]{Ginzburg:2015}
Ginzburg, S. \& Sari, R. 2015, The Astrophysical Journal, 803, 111

\bibitem[{Ginzburg \& Sari(2016)}]{Ginzburg:2015a}
---. 2016, The Astrophysical Journal, 819, 116

\bibitem[{Grunblatt {et~al.}(2016)Grunblatt, Huber, Gaidos, Lopez, Fulton,
  Vanderburg, Barclay, Fortney, Howard, \& Isaacson}]{Grunblatt2016}
Grunblatt, S.~K., Huber, D., Gaidos, E.~J., Lopez, E.~D., Fulton, B.~J.,
  Vanderburg, A., Barclay, T., Fortney, J.~J., Howard, A.~W., \& Isaacson,
  H.~T. 2016, The Astronomical Journal, 152, 1

\bibitem[{Guillot(2010)}]{Guillot:2010}
Guillot, T. 2010, Astronomy and Astrophysics, 520, A27

\bibitem[{Guillot {et~al.}(1996)Guillot, Burrows, Hubbard, Lunine, \&
  Saumon}]{Guillot:1996}
Guillot, T., Burrows, A., Hubbard, W., Lunine, J., \& Saumon, D. 1996, The
  Astrophysical Journal Letters, 459, L35

\bibitem[{Guillot {et~al.}(1994)Guillot, Gautier, Chabrier, \&
  Mosser}]{Guillot:1994a}
Guillot, T., Gautier, D., Chabrier, G., \& Mosser, B. 1994, Icarus, 112, 337

\bibitem[{Guillot \& Showman(2002)}]{Guillot_2002}
Guillot, T. \& Showman, A. 2002, Astronomy and Astrophysics, 385, 156

\bibitem[{Han {et~al.}(2014)Han, Wang, Wright, Feng, Zhao, Fakhouri, Brown, \&
  Hancock}]{Han:2014}
Han, E., Wang, S., Wright, J., Feng, Y., Zhao, M., Fakhouri, O., Brown, J., \&
  Hancock, C. 2014, Publications of the Astronomical Society of the Pacific,
  126, 827

\bibitem[{Hartman {et~al.}(2016)Hartman, Bakos, Bhatti, Penev, Bieryla, Latham,
  Kov{\'{a}}cs, Torres, Csubry, de~Val-Borro, Buchhave, Kov{\'{a}}cs, Quinn,
  Howard, Isaacson, Fulton, Everett, Esquerdo, B{\'{e}}ky, Szklenar, Falco,
  Santerne, Boisse, H{\'{e}}brard, Burrows, L{\'{a}}z{\'{a}}r, Papp, \&
  S{\'{a}}ri}]{Hartman2016}
Hartman, J.~D., Bakos, G.~{\'{A}}., Bhatti, W., Penev, K., Bieryla, A., Latham,
  D.~W., Kov{\'{a}}cs, G., Torres, G., Csubry, Z., de~Val-Borro, M., Buchhave,
  L., Kov{\'{a}}cs, T., Quinn, S., Howard, A.~W., Isaacson, H., Fulton, B.~J.,
  Everett, M.~E., Esquerdo, G., B{\'{e}}ky, B., Szklenar, T., Falco, E.,
  Santerne, A., Boisse, I., H{\'{e}}brard, G., Burrows, A., L{\'{a}}z{\'{a}}r,
  J., Papp, I., \& S{\'{a}}ri, P. 2016, The Astronomical Journal, 152, 182

\bibitem[{Heng(2012)}]{Heng:2012}
Heng, K. 2012, The Astrophysical Journal Letters, 748, L17

\bibitem[{Henry {et~al.}(2000)Henry, Marcy, Butler, \& Vogt}]{Henry:2000}
Henry, G., Marcy, G., Butler, R., \& Vogt, S. 2000, The Astrophysical Journal,
  529, L41

\bibitem[{Henyey {et~al.}(1959)Henyey, Wilets, Bohm, LeLevier, \&
  Levee}]{Henyey:1959}
Henyey, L., Wilets, L., Bohm, K., LeLevier, R., \& Levee, R. 1959, The
  Astrophysical Journal, 129, 628

\bibitem[{Huang \& Cumming(2012)}]{Huang_2012}
Huang, X. \& Cumming, A. 2012, The Astrophysical Journal, 757, 47

\bibitem[{Ibgui \& Burrows(2009)}]{Ibgui:2009}
Ibgui, L. \& Burrows, A. 2009, The Astrophysical Journal, 700, 1921

\bibitem[{Ibgui {et~al.}(2010)Ibgui, Burrows, \& Spiegel}]{Ibgui:2010}
Ibgui, L., Burrows, A., \& Spiegel, D. 2010, The Astrophysical Journal, 713,
  751

\bibitem[{Jackson {et~al.}(2008)Jackson, Greenberg, \& Barnes}]{Jackson:681}
Jackson, B., Greenberg, R., \& Barnes, R. 2008, The Astrophysical Journal, 681,
  1631

\bibitem[{Kataria {et~al.}(2016)Kataria, Sing, Lewis, Visscher, Showman,
  Fortney, \& Marley}]{Kataria2016}
Kataria, T., Sing, D., Lewis, N., Visscher, C., Showman, A., Fortney, J., \&
  Marley, M. 2016, The Astrophysical Journal, 821, 9

\bibitem[{Kippenhahn {et~al.}(2012)Kippenhahn, Weigert, \&
  Weiss}]{Kippenhahn:2012}
Kippenhahn, R., Weigert, A., \& Weiss, A. 2012, Stellar Structure and
  Evolution, 2nd edn. (New York: Springer)

\bibitem[{Komacek \& Showman(2016)}]{Komacek:2015}
Komacek, T. \& Showman, A. 2016, The Astrophysical Journal, 821, 16

\bibitem[{Komacek {et~al.}(2017)Komacek, Showman, \& Tan}]{Komacek:2017}
Komacek, T., Showman, A., \& Tan, X. 2017, The Astrophysical Journal, 835, 198

\bibitem[{Laughlin {et~al.}(2011)Laughlin, Crismani, \& Adams}]{Laughlin_2011}
Laughlin, G., Crismani, M., \& Adams, F. 2011, The Astrophysical Journal
  Letters, 729, L7

\bibitem[{Laughlin \& Lissauer(2015)}]{Laughlin:2015}
Laughlin, G. \& Lissauer, J. Treatise on Geopysics, 2nd edn., ed. G.~Schubert
  (Elsevier)

\bibitem[{Leconte \& Chabrier(2012)}]{Leconte:2012}
Leconte, J. \& Chabrier, G. 2012, Astronomy and Astrophysics, 540, A20

\bibitem[{Leconte {et~al.}(2010)Leconte, Chabrier, Baraffe, \&
  Levrard}]{Leconte:2010a}
Leconte, J., Chabrier, G., Baraffe, I., \& Levrard, B. 2010, Astronomy and
  Astrophysics, 516, A64

\bibitem[{Lopez \& Fortney(2016)}]{Lopez:2015}
Lopez, E. \& Fortney, J. 2016, The Astrophysical Journal, 818, 4

\bibitem[{Mayne {et~al.}(2014)Mayne, Baraffe, Acreman, Smith, Browning,
  Amundsen, Wood, Thuburn, \& Jackson}]{Mayne:2014}
Mayne, N., Baraffe, I., Acreman, D., Smith, C., Browning, M., Amundsen, D.,
  Wood, N., Thuburn, J., \& Jackson, D. 2014, Astronomy and Astrophysics, 561,
  A1

\bibitem[{Menou(2012{\natexlab{a}})}]{Menou:2012fu}
Menou, K. 2012{\natexlab{a}}, The Astrophysical Journal, 745, 138

\bibitem[{Menou(2012{\natexlab{b}})}]{Menou_2012}
---. 2012{\natexlab{b}}, The Astrophysical Journal Letters, 754, L9

\bibitem[{Menou \& Rauscher(2009)}]{Menou:2009}
Menou, K. \& Rauscher, E. 2009, The Astrophysical Journal, 700, 887

\bibitem[{Miller \& Fortney(2011)}]{Miller:2011}
Miller, N. \& Fortney, J. 2011, The Astrophysical Journal Letters, 736, L29

\bibitem[{Miller {et~al.}(2009)Miller, Fortney, \& Jackson}]{Miller:2009}
Miller, N., Fortney, J., \& Jackson, B. 2009, The Astrophysical Journal, 702,
  1413

\bibitem[{Owen \& Wu(2016)}]{Owen:2015}
Owen, J. \& Wu, Y. 2016, The Astrophysical Journal, 817, 107

\bibitem[{Paxton {et~al.}(2011)Paxton, Bildsten, Dotter, Herwig, Lesaffre, \&
  Timmes}]{Paxton:2011}
Paxton, B., Bildsten, L., Dotter, A., Herwig, F., Lesaffre, P., \& Timmes, F.
  2011, The Astrophysical Journal Supplement Series, 192, 3

\bibitem[{Paxton {et~al.}(2013)Paxton, Cantiello, Arras, Bildsten, Brown,
  Dotter, Mankovich, Montgomery, Stello, Timmes, \& Townsend}]{Paxton:2013}
Paxton, B., Cantiello, M., Arras, P., Bildsten, L., Brown, E., Dotter, A.,
  Mankovich, C., Montgomery, M., Stello, D., Timmes, F., \& Townsend, R. 2013,
  The Astrophysical Journal Supplement Series, 208, 4

\bibitem[{Paxton {et~al.}(2015)Paxton, Marchant, Schwab, Bauer, Bildsten,
  Cantiello, Dessart, Farmer, Hu, Langer, Townsend, Townsley, \&
  Timmes}]{Paxton:2015}
Paxton, B., Marchant, P., Schwab, J., Bauer, E., Bildsten, L., Cantiello, M.,
  Dessart, L., Farmer, R., Hu, H., Langer, N., Townsend, R., Townsley, D., \&
  Timmes, F. 2015, The Astrophysical Journal Supplement Series, 220, 15

\bibitem[{Peixoto \& Oort(1992)}]{Peixoto:1992}
Peixoto, J. \& Oort, A. 1992, Physics of Climate (New York: American Institute
  of Physics)

\bibitem[{Perna {et~al.}(2012)Perna, Heng, \& Pont}]{perna_2012}
Perna, R., Heng, K., \& Pont, F. 2012, The Astrophysical Journal, 751, 59

\bibitem[{Perna {et~al.}(2010)Perna, Menou, \& Rauscher}]{Perna_2010_2}
Perna, R., Menou, K., \& Rauscher, E. 2010, The Astrophysical Journal, 724, 313

\bibitem[{Rauscher \& Menou(2010)}]{Rauscher:2010}
Rauscher, E. \& Menou, K. 2010, The Astrophysical Journal, 714, 1334

\bibitem[{Rauscher \& Menou(2013)}]{Rauscher_2013}
---. 2013, The Astrophysical Journal, 764, 103

\bibitem[{Rogers \& Komacek(2014)}]{Rogers:2014}
Rogers, T. \& Komacek, T. 2014, The Astrophysical Journal, 794, 132

\bibitem[{Rogers \& Showman(2014)}]{Rogers:2020}
Rogers, T. \& Showman, A. 2014, The Astrophysical Journal Letters, 782, L4

\bibitem[{Saumon {et~al.}(1995)Saumon, Chabrier, \& Horn}]{Saumon:1995}
Saumon, D., Chabrier, G., \& Horn, H.~V. 1995, The Astrophysical Journal
  Supplement Series, 99, 713

\bibitem[{Schubert \& Mitchell(2013)}]{Schubert:2013}
Schubert, G. \& Mitchell, J. Comparative Climatology of Terrestrial Planets,
  ed. S.~Mackwell, A.~Simon-Miller, J.~Harder, \& M.~Bullock (Tucson, AZ:
  University of Arizona Press)

\bibitem[{Showman {et~al.}(2010)Showman, Cho, \& Menou}]{Showman_2009}
Showman, A., Cho, J., \& Menou, K. Exoplanets, ed. S.~Seager (Tucson, AZ:
  University of Arizona Press)

\bibitem[{Showman {et~al.}(2009)Showman, Fortney, Lian, Marley, Freedman,
  Knutson, \& Charbonneau}]{Showmanetal_2009}
Showman, A., Fortney, J., Lian, Y., Marley, M., Freedman, R., Knutson, H., \&
  Charbonneau, D. 2009, The Astrophysical Journal, 699, 564

\bibitem[{Showman \& Guillot(2002)}]{showman_2002}
Showman, A. \& Guillot, T. 2002, Astronomy and Astrophysics, 385, 166

\bibitem[{Showman \& Polvani(2011)}]{Showman_Polvani_2011}
Showman, A. \& Polvani, L. 2011, The Astrophysical Journal, 738, 71

\bibitem[{Spiegel \& Burrows(2013)}]{Spiegel:2013}
Spiegel, D. \& Burrows, A. 2013, The Astrophysical Journal, 772, 76

\bibitem[{Tremblin {et~al.}(2017)Tremblin, Chabrier, Mayne, Amundsen, Baraffe,
  Debras, Drummond, Manners, \& Fromang}]{Tremblin:2017}
Tremblin, P., Chabrier, G., Mayne, N., Amundsen, D., Baraffe, I., Debras, F.,
  Drummond, B., Manners, J., \& Fromang, S. 2017, The Astrophysical Journal,
  843, 30

\bibitem[{Valsecchi {et~al.}(2015)Valsecchi, Rappaport, Rasio, Marchant, \&
  Rogers}]{Valsecchi:2015}
Valsecchi, F., Rappaport, S., Rasio, F., Marchant, P., \& Rogers, L. 2015, The
  Astrophysical Journal, 813, 101

\bibitem[{Weiss {et~al.}(2013)Weiss, Marcy, Rowe, Howard, Isaacson, Fortney,
  Miller, Demory, Fischer, Adams, Dupree, Howell, Kolbl, Johnson, Horch,
  Everett, Fabrycky, \& Seager}]{Weiss:2013}
Weiss, L., Marcy, G., Rowe, J., Howard, A., Isaacson, H., Fortney, J., Miller,
  N., Demory, B., Fischer, D., Adams, E., Dupree, A., Howell, S., Kolbl, R.,
  Johnson, J., Horch, E., Everett, M., Fabrycky, D., \& Seager, S. 2013, The
  Astrophysical Journal, 768, 14

\bibitem[{Wu \& Lithwick(2013)}]{Wu:2013}
Wu, Y. \& Lithwick, Y. 2013, The Astrophysical Journal, 763, 13

\bibitem[{Youdin \& Mitchell(2010)}]{Youdin_2010}
Youdin, A. \& Mitchell, J. 2010, The Astrophysical Journal, 721, 1113

\end{thebibliography}


\begin{thebibliography}
\end{thebibliography}
\fi
\if\bibinc y

\fi


\end{document}